%% file: main.tex
\documentclass[acmlarge, nonacm]{acmart}

\usepackage{kotex}
\usepackage{comment}
\usepackage{caption}
\usepackage{subcaption}
\usepackage{enumitem}
\usepackage{multirow} 
\usepackage{array}  
\usepackage{xcolor}
\usepackage[ruled,vlined]{algorithm2e}
\usepackage{booktabs}
\usepackage{colortbl}  
\usepackage{tabularx} 
\usepackage{makecell}
\usepackage{soul}

\newbool{showComments}
\booltrue{showComments}
\usepackage{colortbl}
\usepackage{color}

\ifbool{showComments}{%

}

\AtBeginDocument{%
  }

\setcopyright{acmlicensed}
\copyrightyear{2025}
\acmYear{2025}
\acmDOI{XXXXXXX.XXXXXXX}

\acmJournal{IMWUT}
\acmVolume{37}
\acmNumber{4}
\acmArticle{111}
\acmMonth{11}

\begin{document}

\title{Enabling Automatic Self-Talk Detection via Earables}

\author{Euihyeok Lee}
\email{euihyeok.lee@misl.koreatech.ac.kr}
\affiliation{%
  \institution{Korea University of Technology and Education}
  \country{Republic of Korea}
}

\author{Seonghyeon Kim}
\email{seonghyeon.kim@misl.koreatech.ac.kr}
\affiliation{%
  \institution{Korea University of Technology and Education}
  \country{Republic of Korea}
}

\author{SangHun Im}
\email{tkrhkshdqn@koreatech.ac.kr}
\affiliation{%
  \institution{Korea University of Technology and Education}
  \country{Republic of Korea}
}

\author{Heung-Seon Oh}
\email{ohhs@koreatech.ac.kr}
\affiliation{%
  \institution{Korea University of Technology and Education}
  \country{Republic of Korea}
}

\author{Seungwoo Kang}
\authornote{Corresponding author}
\email{swkang@koreatech.ac.kr}
\affiliation{%
  \institution{Korea University of Technology and Education}
  \country{Republic of Korea}
}

\renewcommand{\shortauthors}{Lee et al.}

\input{sections/00.Abstract}


\settopmatter{printacmref=false}
\renewcommand\footnotetextcopyrightpermission[1]{}
\pagestyle{plain}


\maketitle

\input{sections/01.Introduction}
\input{sections/02.Background_and_Motivation}
\input{sections/03.MutterMeter_Overview}
\input{sections/04.Self-talk_Sensing}
\input{sections/05.DataCollection}

\input{sections/06.Evaluation}

\input{sections/07.Related_Work}

\input{sections/08.Conclusion}

\bibliographystyle{ACM-Reference-Format}
\bibliography{reference}

\appendix
\input{sections/A.Appendix}

\end{document}

%% file: sections/00.Abstract.tex
\begin{abstract}

Self-talk---an internal dialogue that can occur silently or be spoken aloud---plays a crucial role in emotional regulation, cognitive processing, and motivation, yet has remained largely invisible and unmeasurable in everyday life. In this paper, we present MutterMeter, a mobile system that automatically detects vocalized self-talk from audio captured by earable microphones in real-world settings. Detecting self-talk is technically challenging due to its diverse acoustic forms, semantic and grammatical incompleteness, and irregular occurrence patterns, which differ fundamentally from assumptions underlying conventional speech understanding models. To address these challenges, MutterMeter employs a hierarchical classification architecture that progressively integrates acoustic, linguistic, and contextual information through a sequential processing pipeline, adaptively balancing accuracy and computational efficiency. We build and evaluate MutterMeter using a first-of-its-kind dataset comprising 31.1 hours of audio collected from 25 participants. Experimental results demonstrate that MutterMeter achieves robust performance with a macro-averaged $F_{1}$ score of 0.84, outperforming conventional approaches, including LLM-based and speech emotion recognition models. 

\end{abstract}

%% file: sections/01.Introduction.tex
\section{Introduction}

Humans frequently engage in self-talk—an internal dialogue that can occur silently or be spoken aloud~\cite{diaz1992private, kohlberg1968private}. For instance, people often utter phrases such as "Ah, I messed up again" or "I can do this" unconsciously in daily life. Such self-talk is not merely a habitual behavior but a part of psychological and cognitive processes. Self-talk reflects one's inner emotions, motivations, and thought patterns, and serves as an important cue for understanding a person's psychological state~\cite{beck1979cognitive, hatzigeorgiadis2020strategic}. It also functions as a spontaneous mechanism for emotional regulation and cognitive processing, helping individuals manage tension and regulate thoughts. Accordingly, self-talk tends to occur more frequently in situations that require sustained concentration or repeated decision-making—such as during sports, driving, or learning. In such contexts, where external interaction is limited, individuals are more likely to rely on internal dialogue to cope with psychological pressure. 

Numerous studies in sports and cognitive psychology have demonstrated that self-talk plays a critical role in performance enhancement, emotional regulation, and motivation~\cite{tod2011effects, latinjak2016goal, olisola2021influence, neil2013seeing}. However, most existing studies have relied on self-reports or post-event interviews to analyze the types and circumstances of self-talk, which limits their ability to objectively capture its actual frequency and contextual patterns~\cite{sanchez2016self, damirchi2020role, wang2017achievement}. Some studies have also reported inconsistencies between self-reports and external observations, indicating that individuals often have difficulty accurately recognizing or recalling when and how their self-talk occurs~\cite{thibodeaux2018selftalk, brinthaupt2023self}.  
This is mainly because self-talk typically occurs automatically and spontaneously, making it challenging to capture reliably through self-reports. While third-person observation or video analysis can provide more reliable data, these approaches require substantial time and effort, making them impractical for real-world or large-scale deployment.

In this paper, we explore a technical approach to automatically detect \textit{vocalized self-talk} that occurs spontaneously in everyday contexts. As an initial observational case, we focus on individual tennis play, a representative activity where self-talk frequently emerges. Tennis offers a suitable setting for this study, as it involves reduced external interaction, frequent decision-making, and emotionally charged moments that naturally elicit self-talk. To this end, we present \textit{MutterMeter}, a mobile self-talk detection system that analyzes audio from earable microphones on the fly and classifies utterances into three categories: positive self-talk, negative self-talk, and others (e.g., conversations). To the best of our knowledge, MutterMeter is the first mobile system to automatically detect self-talk, until now invisible and unmeasurable, using mobile sensing technologies. We believe that it can provide a basis for numerous practical applications to precisely understand individuals’ psychological states and support emotional regulation and mental health management in daily life.

Although existing speech understanding technologies—such as Speech Emotion Recognition (SER), Spoken Language Understanding (SLU), and sentiment analysis—have achieved remarkable progress in recognizing emotions and interpreting spoken content, research on self-talk classification remains limited. Our observations indicate that detecting self-talk poses distinct challenges due to its unique characteristics. First, self-talk exhibits a high degree of acoustic variability. In some cases, it appears as strong emotional expressions; in others, it is spoken in a very soft or even barely audible voice. Such variability makes it difficult to reliably classify or model self-talk using acoustic features alone. Second, self-talk often shows semantic and grammatical incompleteness. As it is not intended for listeners, many utterances are syntactically incomplete or consist only of brief emotional expressions or interjections, introducing ambiguity in text-based interpretation and classification. Third, self-talk occurs sporadically and irregularly. It intermittently emerges from internal thought processes. Although more frequent under stress or emotional arousal, its occurrence remains inconsistent. These characteristics conflict with the assumptions of conventional models. Most SER and SLU systems are designed for structured or interaction-oriented speech data, assuming clear articulation, grammatical completeness, and stable vocal intensity. The datasets, such as IEMOCAP or RAVDESS, typically contain acted or structured speech in controlled, interactive settings. In contrast, daily self-talk rarely satisfies these assumptions; it often appears in ambiguous, irregular, and unstructured forms, making it inherently difficult to accurately detect with existing models.

To address the challenges, MutterMeter employs a hierarchical classification-based approach that integrates acoustic, linguistic, and contextual information via a sequential processing pipeline, balancing accuracy and efficiency in self-talk detection. The pipeline comprises the acoustic, linguistic, and fusion stages, in which contextual information is progressively incorporated to refine understanding and improve classification accuracy. The acoustic stage performs lightweight, on-device inference from audio captured by earable microphones, detecting potential self-talk segments based on acoustic characteristics. When additional semantic or contextual interpretation is required, the process transitions to the linguistic stage, which transcribes the utterance segments and extracts linguistic patterns for further classification. Finally, the fusion stage combines both modalities, enabling the system to handle ambiguous or low-confidence cases that cannot be resolved by a single modality. By evaluating classification confidence at each stage, MutterMeter adaptively determines the necessary depth of processing, balancing accuracy with computational efficiency through adaptive stage transitions.

To build and evaluate MutterMeter, we collected a first-of-its-kind self-talk dataset captured in real tennis-playing environments, comprising approximately 31.1 hours of audio recordings from 25 participants; most were recreational players who regularly play tennis for leisure, and some were amateur tennis players.
It contains various utterances, including self-talk, with manually labeled, accurate annotations.
Our evaluation using the dataset shows that MutterMeter accurately detects self-talk even for unseen users, achieving a macro-averaged $F_{1}$ score of 0.84. In particular, by leveraging the temporal and contextual information of utterances, MutterMeter extracts more representative embeddings, enabling more accurate self-talk detection. More importantly, its hierarchical self-talk detection model consistently surpasses all comparison models, demonstrating higher classification accuracy. Furthermore, system cost analysis reveals that the model reduces average latency per utterance by 41\%. In addition, it outperforms alternatives, including LLMs, SER, and sentiment analysis, by a substantial margin.

We summarize the contributions of this paper as follows. First, we introduce self-talk—a self-directed form of speech—as a novel research problem in the domain of mobile and ubiquitous computing. By extending the scope of mobile sensing to self-talk, which reflects a speaker's internal states and cognitive processes, we open a new avenue for the computational understanding of human psychological and emotional experiences. Second, we collect an in-the-wild dataset of naturally occurring self-talk captured during real tennis play. Using this dataset, we design and implement MutterMeter, a mobile system that incorporates a hierarchical classification architecture integrating acoustic, linguistic, and contextual information for accurate and efficient self-talk detection. Third, we conduct an extensive evaluation on the collected dataset, demonstrating that MutterMeter achieves both high accuracy and computational efficiency compared to existing approaches, and validating its feasibility for real-time self-talk detection in real-world environments.

%% file: sections/02.Background_and_Motivation.tex
\section{Background and Motivation}~\label{sec:background_and_motivation}

\subsection{What is self-talk?}

Self-talk refers to communication with oneself, which can manifest as internal monologues or overt verbalizations~\cite{diaz1992private, kohlberg1968private}. Although people frequently engage in self-talk in their daily lives, they are not always consciously aware of its occurrence~\cite{latinjak2023self}. Self-talk plays an important role in both cognitive and emotional processes, being closely related to various psychological functions that influence one’s feelings, thoughts, and behaviors~\cite{beck1979cognitive, hatzigeorgiadis2020strategic}. Given these characteristics, self-talk has attracted considerable research interest and has been extensively explored across diverse domains, including the treatment of clinical symptoms~\cite{beck1979cognitive, heimberg1989cognitive,twamley2012compensatory}, the promotion of personal well-being~\cite{chan2019brief}, and the enhancement of learning~\cite{wang2017achievement, sanchez2016self} and sports~\cite{latinjak2016goal} performance.

In particular, within the field of sports psychology, the positive effects of self-talk on athletes’ performance have been actively explored. Many studies have reported that self-talk could contribute to enhanced focus and emotional regulation, thereby providing practical benefits for performance improvement~\cite{tod2011effects, latinjak2016goal}. Conversely, several studies have shown that negative self-talk is significantly associated with decreased performance and emotional anxiety, potentially undermining athletes’ psychological stability~\cite{santos2022positive, borrajo2024negative}. Moreover, in individual sports such as tennis or golf, where immediate external feedback is limited, self-talk tends to occur more frequently, and its patterns in these contexts can have a significant impact on performance~\cite{olisola2021influence, neil2013seeing}.

In the mental health domain, self-talk has been studied both as an indicator of symptoms and as a tool for intervention. Specifically, negative self-talk is closely associated with depression, anxiety, and low self-esteem, and can serve as a key cue for understanding an individual’s cognitive distortions and emotional difficulties~\cite{treadwell1996self}. Several studies have proposed cognitive-behavioral therapy (CBT) techniques that identify patterns of self-talk in patients with mental disorders and train them to restructure or modify these patterns~\cite{farrell1998cognitive, taylor2011effect}.

In daily life situations, self-talk has been reported to serve various cognitive and emotional functions. People who live alone tend to engage in self-talk more frequently, and such self-talk often reflects their underlying psychological state~\cite{reichl2013relation}. Some studies have also explored the role of self-talk in specific daily contexts, such as during travel~\cite{pasternak2024self}, studying~\cite{sanchez2016self, wang2017achievement}, or light physical activity~\cite{cousins2005just, gibson2007role}.

Given the impact of self-talk on individuals' emotions, cognition, and behavior, we envision that the automatic detection, quantification, and analysis of self-talk presents a significant opportunity to lay a foundation for a more precise understanding and support of individual states. Automatic detection and analysis of self-talk will open up a wide range of practical applications across various domains, including mental health monitoring and personalized coaching. Importantly, it can enable novel forms of intervention and feedback for ordinary users in everyday contexts, where self-talk has been invisible and unmeasurable, thereby transforming how individuals understand and manage their internal experiences.

\textbf{Research scope:} This study primarily focuses on typical users in everyday contexts, rather than individuals in professional or clinical settings (e.g., athletes or patients undergoing therapy). Accordingly, we target naturally occurring forms of self-talk (hereafter, organic self-talk), including spontaneous and goal-directed self-talk. These are distinct from strategic self-talk, which involves pre-defined or deliberately trained verbal expressions in sports and therapeutic settings. Moreover, since MutterMeter relies on audio signals captured through off-the-shelf earable devices, it exclusively detects externally vocalized self-talk, leaving purely internal (non-vocalized) forms beyond its scope. To clarify the conceptual background of this scope, we next describe the major types of self-talk and their defining characteristics.

In general, self-talk can be categorized into organic and strategic forms~\cite{latinjak2023self}. Organic self-talk refers to spontaneous and naturally occurring speech, which constitutes most of what people typically perceive as self-talk. It can be further divided into spontaneous and goal-directed self-talk. Spontaneous self-talk arises impulsively and often reflects momentary emotional or psychological states (e.g., excitement, frustration), for instance, “Wow, that cloud looks beautiful!”  It can be further characterized as positive or negative depending on its emotional tone (e.g., confidence vs. anxiety). In contrast, goal-directed self-talk serves purposeful functions, such as planning or self-regulation, for instance, “I need to mow the lawn this weekend.” Strategic self-talk, on the other hand, is systematically employed as part of structured psychological interventions or performance-enhancement programs. It is divided into instructional and motivational forms~\cite{latinjak2016goal}. These types are intentionally used to guide performance or foster positive emotions, and their structured, intentional nature distinguishes them from naturally occurring self-talk that our work aims to detect.

To enable a more fine-grained analysis of self-talk, we further differentiate spontaneous self-talk by emotional valence. For instance, negative self-talk may reflect self-critical or anxious internal dialogue, while positive self-talk may indicate confidence, motivation, or planning. Although the present work focuses on detection and classification, this distinction provides a basis for future applications that support self-awareness and emotional regulation in daily life. Accordingly, we define the classification objective as categorizing self-talk into the following three classes:
\begin{itemize}
    \item \textbf{Negative self-talk}: Negative spontaneous self-talk
    \item \textbf{Positive Self-talk}: Positive spontaneous self-talk and goal-directed self-talk
    \item \textbf{Others}: Conversations with others, external noise, etc.
\end{itemize}

\subsection{Preliminary exploration}~\label{sec:preliminary_exploration}

\subsubsection{Acoustic characteristics}~\label{sec:acoustic_characteristics}
Detecting self-talk from earable audio signals can be achieved through acoustic feature analysis, which offers a straightforward approach. Previous studies have demonstrated that various acoustic characteristics, such as loudness, intonation, timbre, and speech rate, serve as useful cues for understanding a speaker’s intention or emotional state~\cite{rao2006prosody, ververidis2006emotional, dellaert1996recognizing}. Informed by these prior studies, we conducted a preliminary exploration of the opportunities and challenges of using acoustic features to detect self-talk. In addition, we explored the feasibility of applying an existing method to this task. The exploration was conducted using our in-the-wild dataset, as described in Section~\ref{sec:data_collection}.

\begin{figure}[t]
    \centering
    \begin{minipage}[b]{0.39\textwidth} 
        \centering
        \begin{subfigure}[b]{0.49\textwidth}
            \centering
            \includegraphics[width=\textwidth]{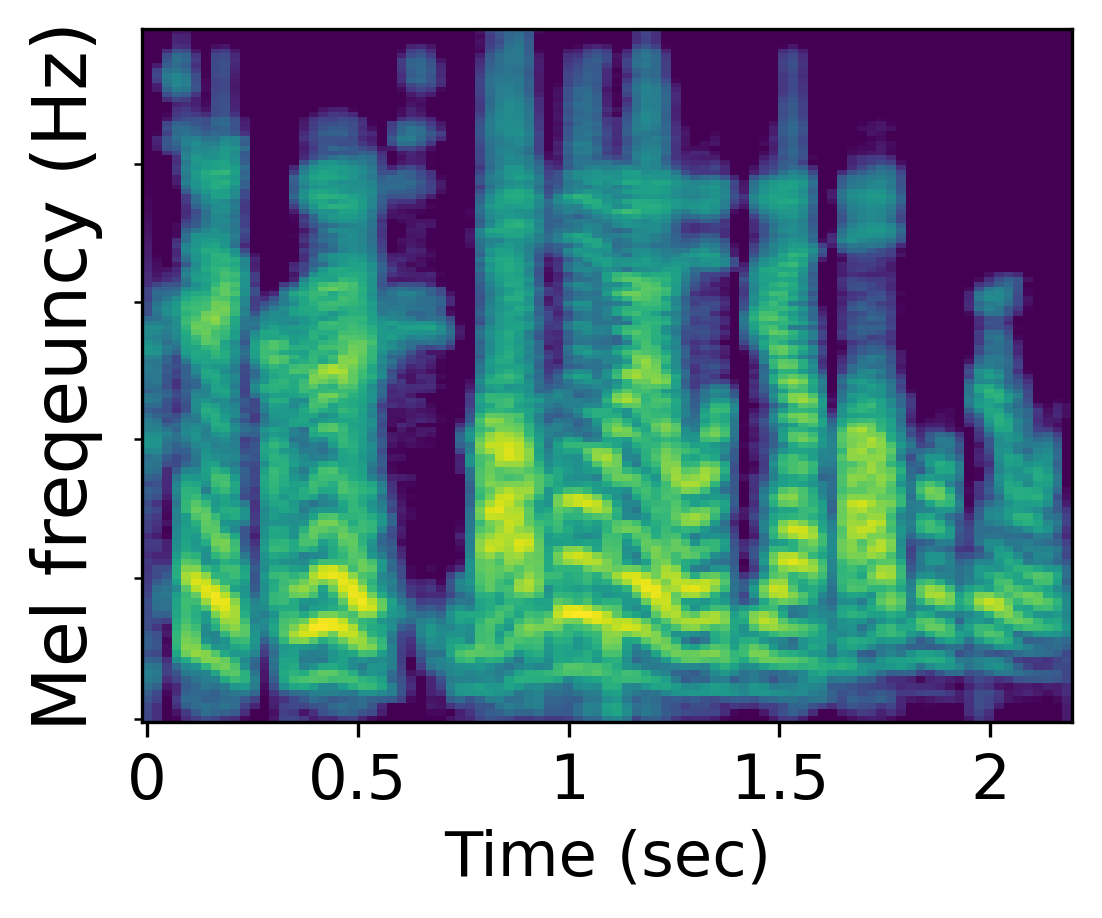}
            \caption{Talk to the other}
            \vspace{-0.05in}
            \label{fig:others1}
        \end{subfigure}
        \hfill
        \begin{subfigure}[b]{0.49\textwidth}
            \centering
            \includegraphics[width=\textwidth]{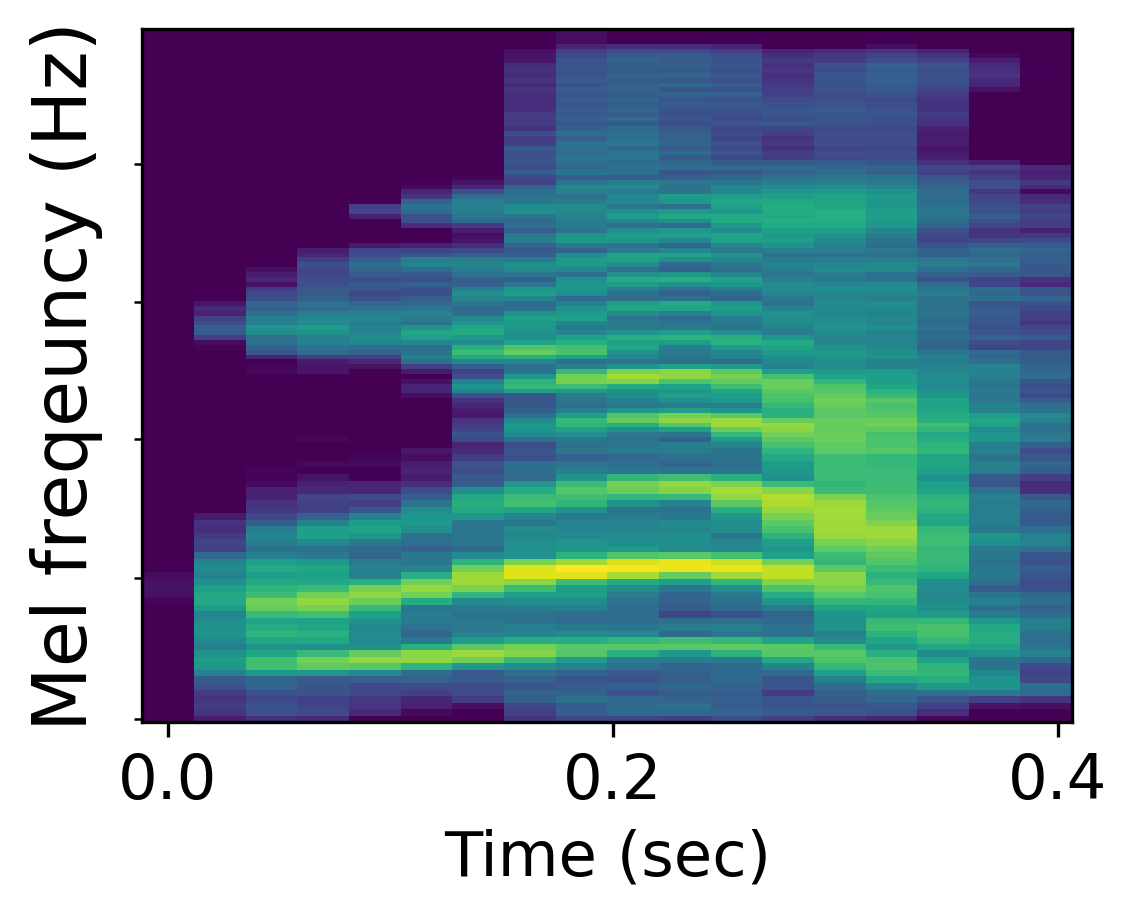}
            \caption{Keeping scores}
            \vspace{-0.05in}
            \label{fig:scoring}
        \end{subfigure}
        \vspace{-0.1in}
        \caption{Non–self-talk examples}
        \label{fig:fig1}
    \end{minipage}
    \hfill
    \begin{minipage}[b]{0.585\textwidth}
        \centering
        \begin{subfigure}[b]{0.32\textwidth}
            \centering
            \includegraphics[width=\linewidth]{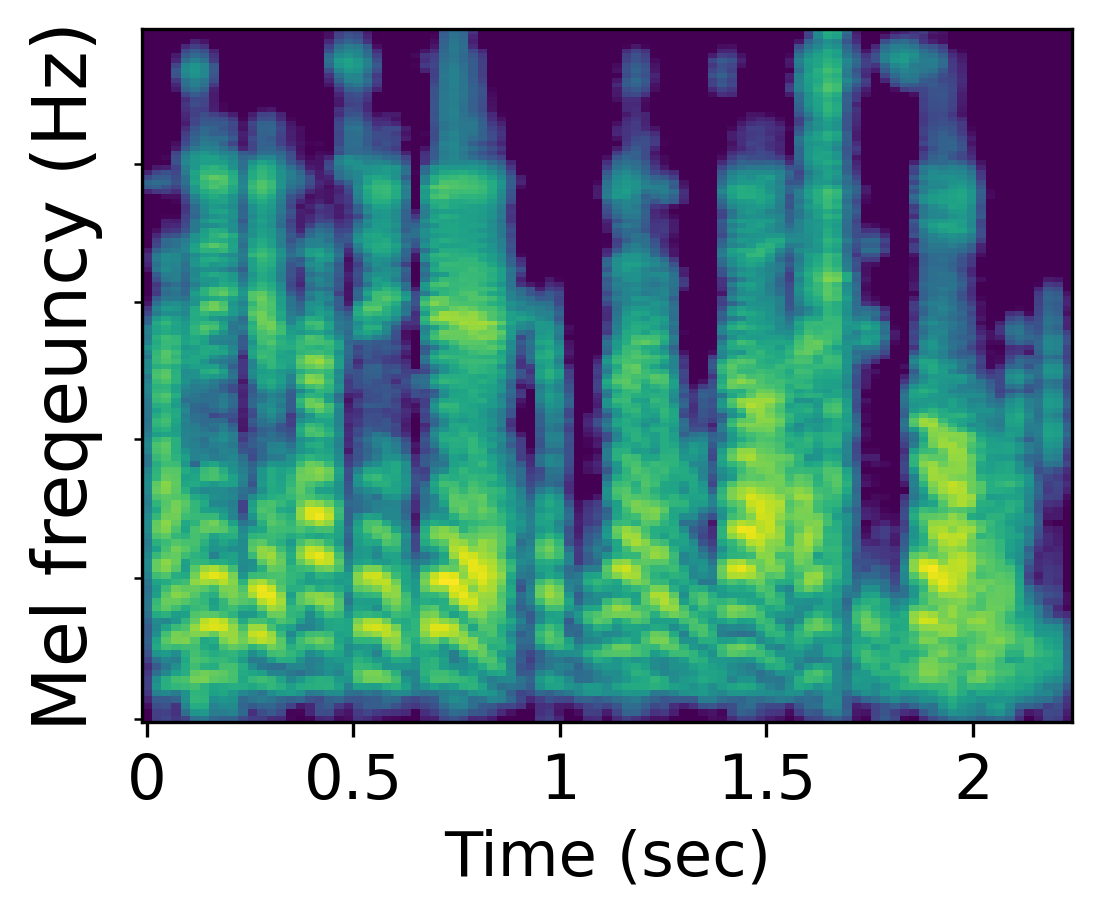}
            \caption{Positive self-talk}
            \vspace{-0.05in}
            \label{fig:goal-others}
        \end{subfigure}
        \hfill
        \begin{subfigure}[b]{0.32\textwidth}
            \centering
            \includegraphics[width=\linewidth]{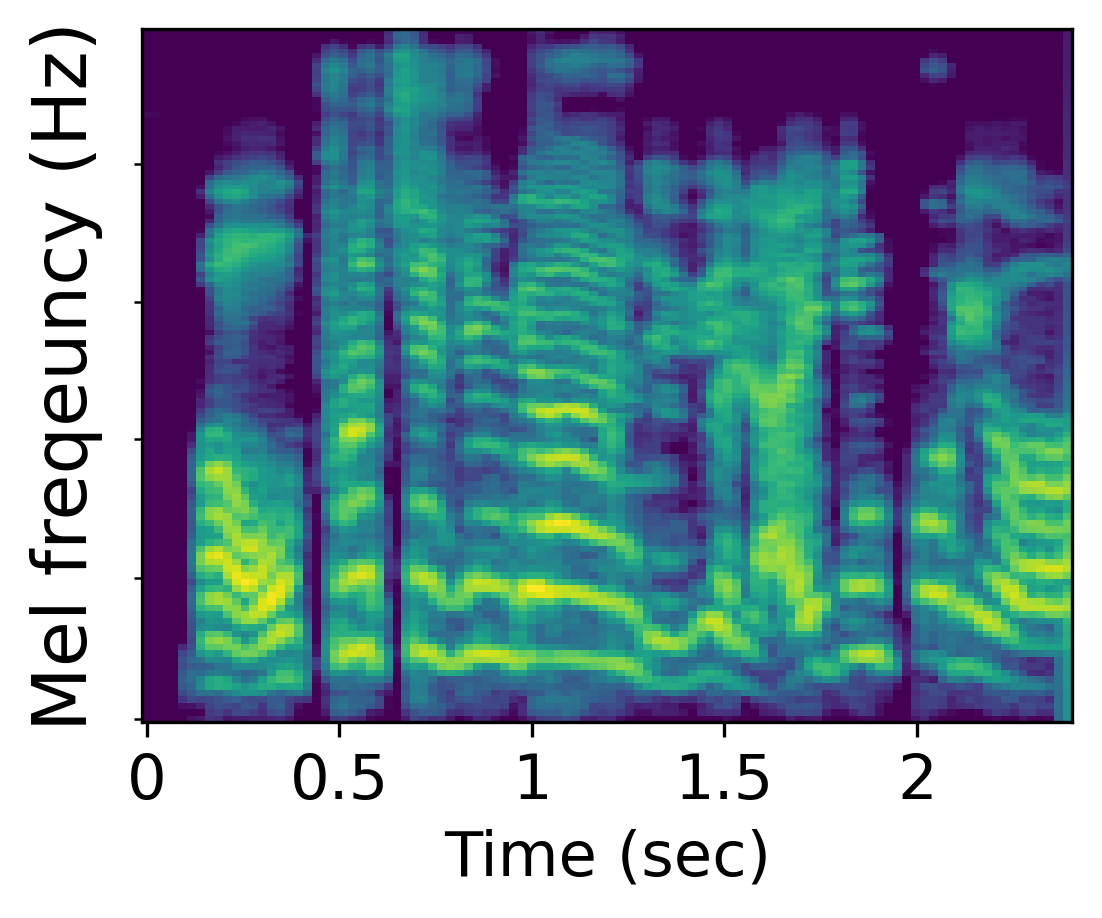}
            \caption{Negative self-talk}
            \vspace{-0.05in}
            \label{fig:neg_other}
        \end{subfigure}
        \hfill
        \begin{subfigure}[b]{0.32\textwidth}
            \centering
            \includegraphics[width=\linewidth]{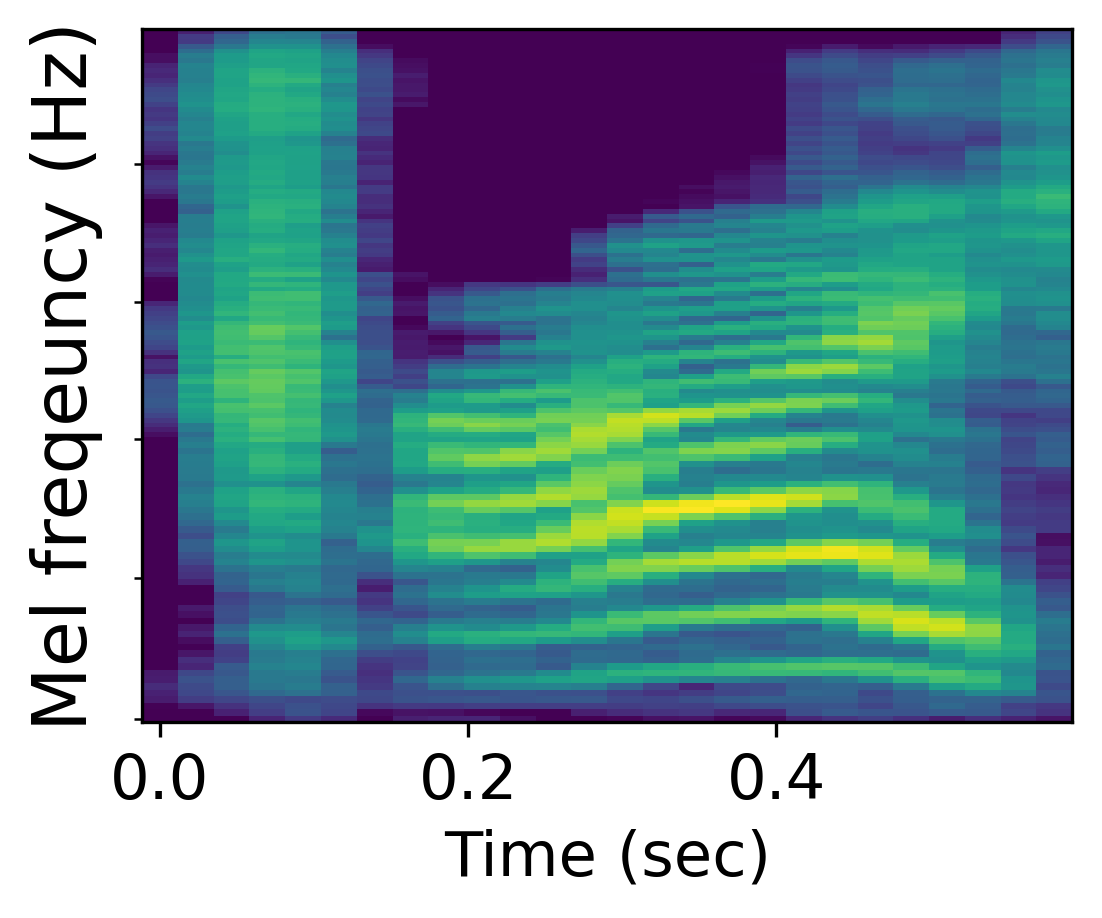}
            \caption{Interjection}
            \vspace{-0.05in}
            \label{fig:neg-short}
        \end{subfigure}
        \vspace{0.05in}
        \caption{Indistinguishable self-talk examples}
        \label{fig:various_utterances_hard}
    \end{minipage}
\end{figure}

\begin{figure}[t]
    \centering
    \begin{minipage}[b]{0.41\textwidth} 
        \centering
        \begin{subfigure}[b]{0.49\textwidth}
            \centering
            \includegraphics[width=\textwidth]{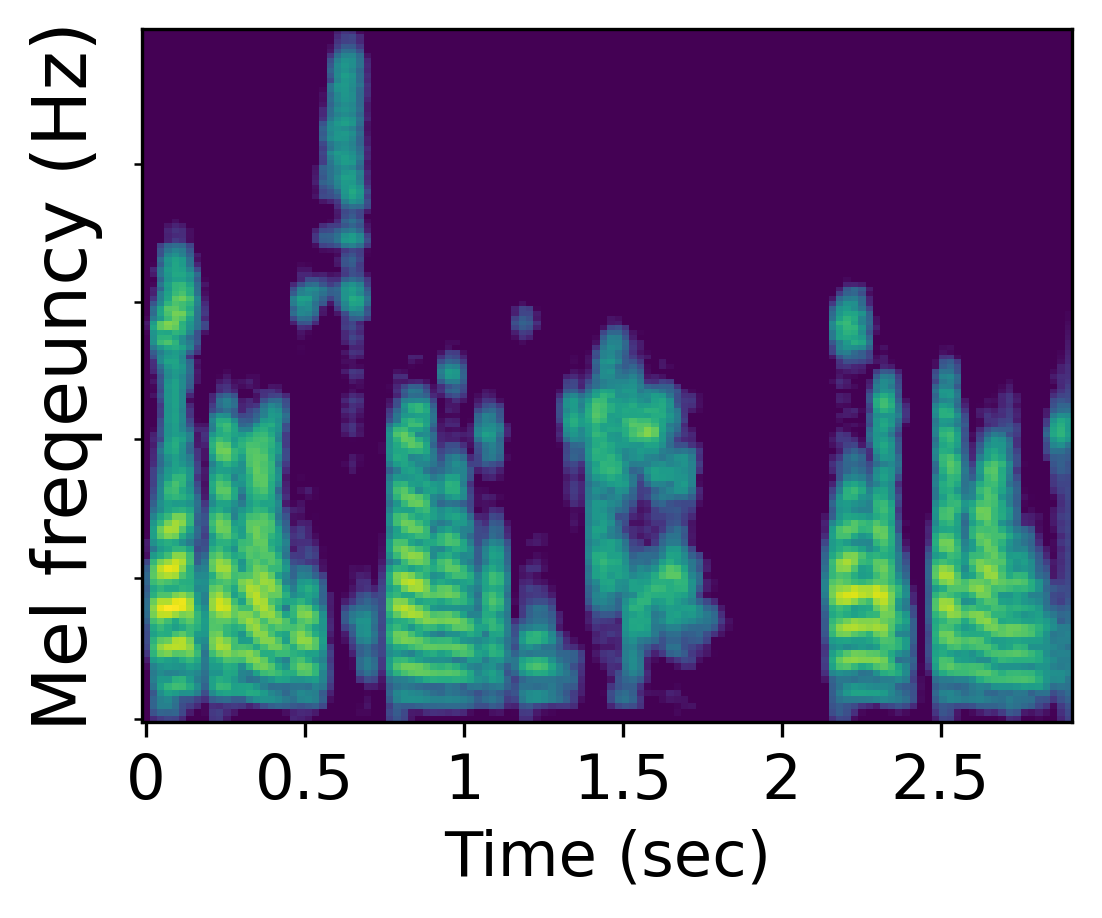}
            \caption{Mumbling self-talk}
            \vspace{-0.05in}
            \label{fig:small-goal}
        \end{subfigure}
        \hfill
        \begin{subfigure}[b]{0.49\textwidth}
            \centering
            \includegraphics[width=\textwidth]{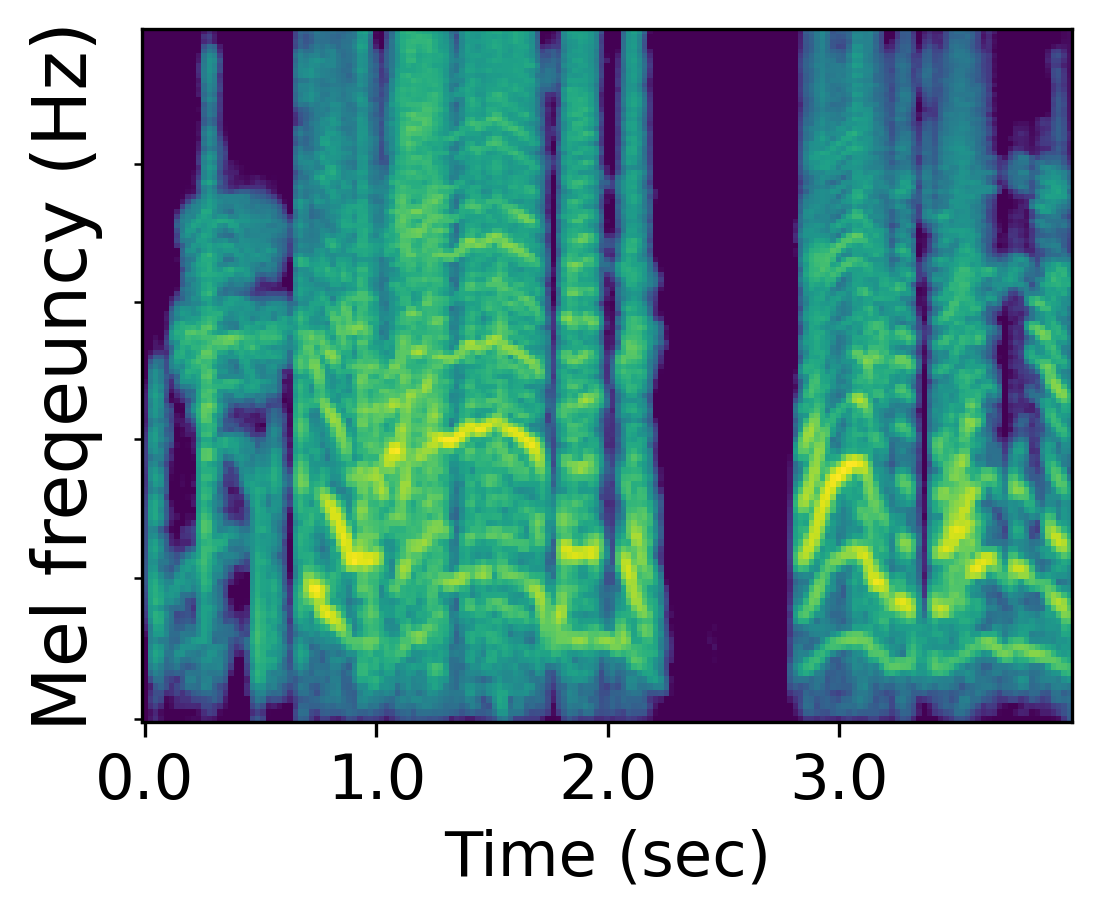}
            \caption{Intense\&loud self-talk}
            \vspace{-0.05in}
            \label{fig:positive_high}
        \end{subfigure}
        \vspace{-0.1in}
        \caption{Distinguishable self-talk examples}
        \label{fig:various_utterances_easy}
    \end{minipage}
    \hfill
    \begin{minipage}[b]{0.57\textwidth}
        \centering
        \begin{subfigure}[b]{0.49\textwidth}
            \centering
            \includegraphics[width=\textwidth]{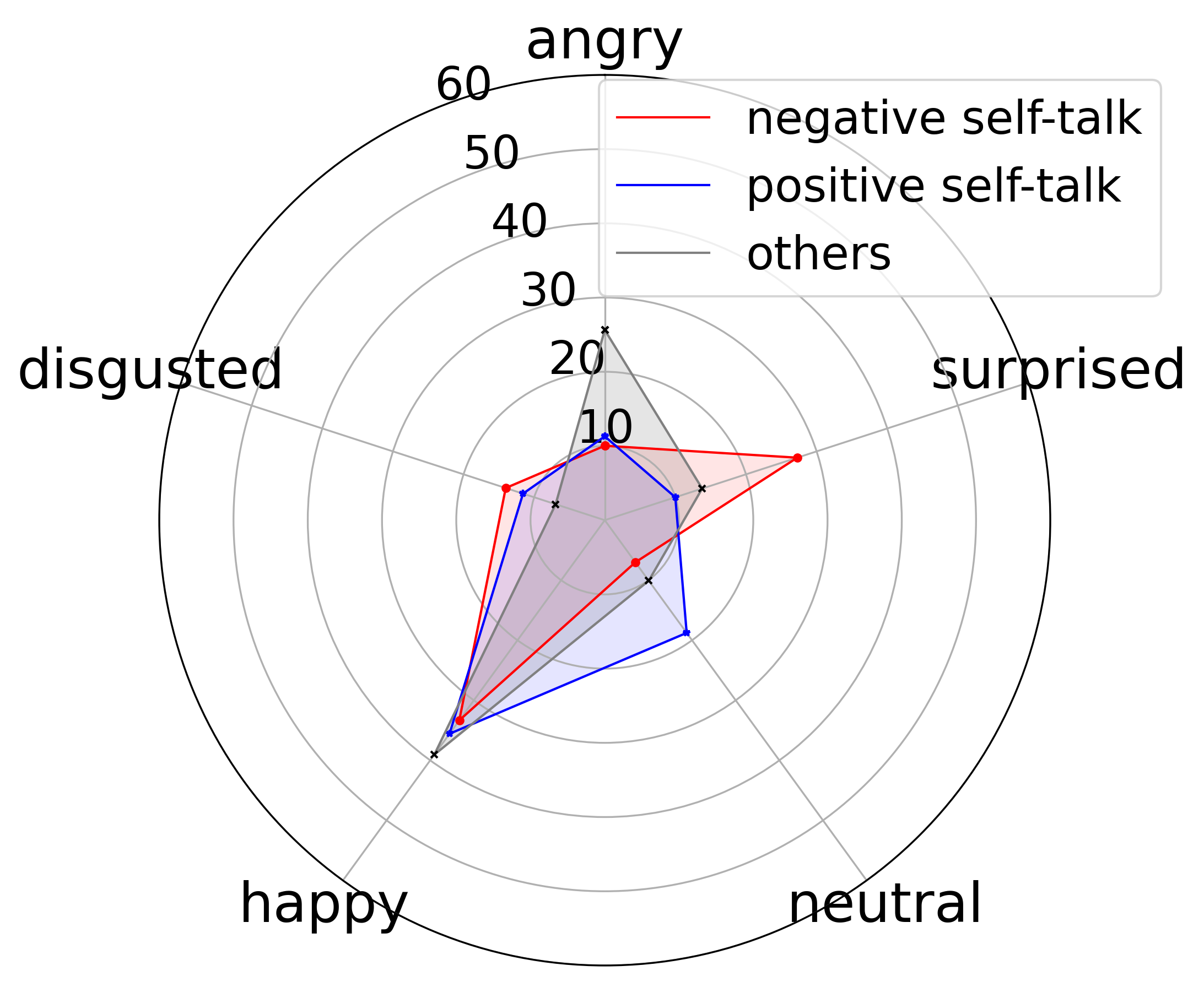}
            \caption{Speech emotion recognition}
            \vspace{-0.05in}
            \label{fig:emotion_recognition}
        \end{subfigure}
        \hfill
        \begin{subfigure}[b]{0.47\textwidth}
            \centering
            \includegraphics[width=\textwidth]{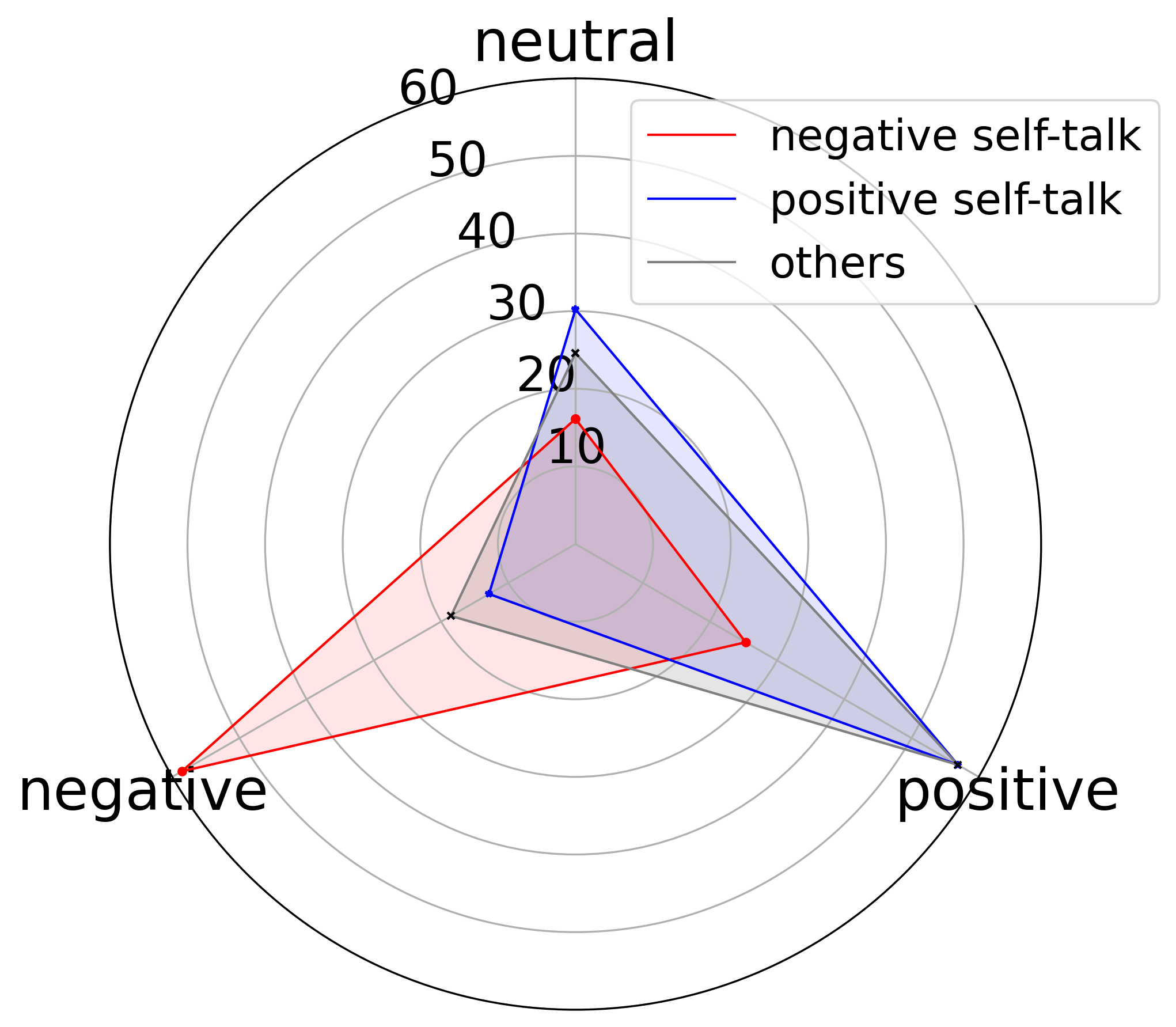}
            \caption{Sentiment analysis}
            \vspace{-0.05in}
            \label{fig:sentiment_analysis}
        \end{subfigure}
        \vspace{-0.1in}
        \caption{Results of self-talk detection using existing methods}
        \label{fig:exsiting_works}
    \end{minipage}
\end{figure}

\textbf{Opportunities and challenges: }
Self-talk exhibits substantial intra-class variability in acoustic features, indicating greater heterogeneity than non-self-talk. 
Generally, non-self-talk is produced with the purpose of communicating with others, maintaining an appropriate pitch, rate, and volume so that the listener can clearly understand the speech (Figure~\ref{fig:others1}). In contrast, self-talk often occurs unconsciously and is self-directed, manifesting as low-volume murmuring (Figure~\ref{fig:small-goal}) or as emotionally charged utterances with high prosodic variation, such as irritation or excitement (Figure~\ref{fig:positive_high}).
Moreover, self-talk may occur more frequently in stressful or emotionally heightened situations, often accompanied by nonverbal cues such as sighs, laughter, or voice tremors. These expressions are among the key characteristics of self-talk and may serve as meaningful acoustic cues.

However, the high intra-class variation in self-talk leads to overlaps with non-self-talk utterances. In particular, some self-talk shares acoustic characteristics with speech directed toward others. 
For example, in goal-directed and negative self-talk, speakers often use intonation and prosody mirroring speech to others (Figures~\ref{fig:goal-others}, \ref{fig:neg_other}), making these utterances acoustically similar to typical non-self-talk (Figure~\ref{fig:others1}).
Furthermore, certain non-self-talk, such as short expressions for announcing scores or indicating in/out calls during games, can acoustically resemble affective self-talk expressions (see Figure~\ref{fig:scoring}, \ref{fig:neg-short}).

\textbf{Limitation of an existing method: }
We examined the feasibility of detecting self-talk using an existing speech emotion recognition (SER) method, which showed clear limitations. Figure~\ref{fig:emotion_recognition} shows the detection results obtained using emotion2vec~\cite{ma-etal-2024-emotion2vec}, one of the representative SER methods. Note that we selected a pretrained model released online by recent studies. Overall, the emotion categories produced by the SER method do not map well to the self-talk labels we defined. For instance, negative self-talk is not consistently classified as a negative emotion, and positive self-talk does not correspond to positive emotions. Specifically, approximately 40\% of the utterances across all self-talk labels are classified as \textit{happy}. For the remaining utterances, negative self-talk, positive self-talk, and others were each frequently classified as \textit{surprised}, \textit{neutral}, and \textit{angry}, respectively.

\subsubsection{Linguistic characteristics}~\label{sec:linguistic_characteristics}
Linguistic features derived from transcribed speech provide informative cues that enable the detection and interpretation of self-talk beyond acoustic information.
Previous studies have demonstrated that text can be used to capture emotions and intentions through sentence structure, grammar, and semantic information~\cite{nandwani2021review, agrawal2012unsupervised}. To further investigate this potential, we examined the opportunities and challenges of using linguistic features for self-talk detection. We also conducted a preliminary study to assess the feasibility of applying an existing method to this task; we used the aforementioned dataset, as described in Section~\ref{sec:data_collection}.

\textbf{Opportunities and challenges: }
Self-talk exhibits distinctive linguistic characteristics that distinguish it from non-self-talk. According to previous linguistic studies, specific endings are frequently used in self-talk, and grammatical structures also differ~\cite{smith2023interactional, ritter2021grammar}. Our observations from the collected data also reveal a similar linguistic pattern. For instance, self-talk often consists of grammatically incomplete or truncated sentences, a tendency likely stemming from the speaker’s implicit assumption that they themselves are the recipient of the utterance. In addition, self-talk frequently involves the repetition of specific words or phrases, which often serves as a strategic means of self-reinforcement or emotion regulation. Examples include utterances such as “Let’s do this, let’s do this,” “Come on, come on,” or “Move, move.” These linguistic features of self-talk arise not from conversational intent but from its psychological functions. They represent a unique linguistic pattern that can serve as a meaningful cue for detecting self-talk.

However, utilizing linguistic features for self-talk detection also entails several challenges. Although self-talk and non-self-talk generally occur in distinct contexts and serve different purposes, they can sometimes exhibit similar surface forms in terms of sentence structure or lexical choice. In such cases, approaches that rely solely on linguistic features face inherent limitations in accurately detecting self-talk. For example, the utterance “What should I do now?” can be self-talk, in which the speaker reflects on their next action, whereas a similar expression, “What should I do?” may be directed to another person as a question in conversation. Likewise, a simple expression like “Okay!” may function as self-talk conveying self-satisfaction, or as a conversational response expressing agreement with another person.

While linguistic features provide rich semantic information, extracting them from audio signals incurs computational overhead and potential sources of error. Linguistic feature extraction, which relies on an additional transcription step, can be computationally intensive and heavily dependent on the quality of Automatic Speech Recognition (ASR). However, obtaining reliable transcription typically requires substantial computational resources, making high-end ASR models impractical for on-device deployment. 
Moreover, the characteristics of self-talk—such as low volume, fast speech rate, and unclear articulation—often lead to transcription errors. These errors may distort the intended meaning of the utterance or cause the loss of important linguistic cues, ultimately affecting the accuracy and reliability of self-talk detection.

\textbf{Limitation of an existing method: }
In our preliminary study, existing sentiment analysis techniques showed limited effectiveness in detecting self-talk. Specifically, we employed TweetNLP~\cite{camacho-collados-etal-2022-tweetnlp} to analyze each utterance. The audio data were first transcribed using the Whisper~\cite{radford2023robust} ASR model, after which sentiment labels were obtained and mapped to the corresponding self-talk labels. Note that we used one of the recent models with publicly available implementations. Interestingly, as shown in Figure~\ref{fig:sentiment_analysis}, sentiment analysis struggled to distinguish between positive self-talk and other utterances: both classes were frequently classified by the model as \textit{neutral} or \textit{positive}. Moreover, although approximately 60\% of negative self-talk samples were correctly identified as \textit{negative}, the remaining 40\% were misclassified. These results suggest that self-talk exhibits distinct linguistic and affective properties compared to typical emotional speech.

\subsection{Motivation}
Our preliminary exploration demonstrates the potential to detect self-talk to a certain extent using acoustic and linguistic features. However, we also identify several limitations, as intrinsic characteristics of self-talk make accurate detection challenging when relying solely on these features of individual utterances. We further observe that existing approaches exhibit limitations in accurately detecting self-talk. To address these limitations, we focus on the idea that self-talk is not a single, isolated vocal event but a cognitive and behavioral process unfolding within specific internal motivations and situational contexts. From this perspective, we seek to leverage the temporal and contextual dynamics of utterances to capture how they relate to each other and to the situational contexts in which they occur, represented as \textit{short-term contextual continuity} and \textit{macro-contextual dependency}.

\subsubsection{Micro-level: short-term contextual continuity}

In natural speech flow, temporally adjacent utterances tend to share similar affective or functional patterns~\cite{Koval2016}, forming a consistent short-term contextual continuity. Our observations indicate that this phenomenon also appears across various types of utterances, including both self-talk and non-self-talk. In particular, when the temporal gap between utterances is short, transitions across different classes occur less frequently, while utterances of the same type tend to appear consecutively. For example, after producing a negative self-talk utterance, a user is more likely to exhibit another utterance that shares a similar emotional state or communicative intent. Such short-term continuity represents a higher-level temporal pattern that is independent of individual acoustic or linguistic features. This temporal relationship provides contextual cues that the model is difficult to capture from isolated utterances.

\subsubsection{Macro-level: macro-contextual dependency}

At a broader level, utterances are strongly influenced by higher-level macro-contexts, such as the user’s ongoing activity, situation, or topic. For instance, when a user is playing tennis, utterances are likely to revolve around that activity. In such contexts, even self-talk often focuses on performance evaluation or self-regulation, while conversations with others primarily concern the flow of the game, rules, or technical aspects. Acts such as counting scores or making in/out calls also naturally arise in this situational context. These higher-level contextual cues serve as valuable grounds for more accurate interpretation of individual utterances. For example, when ASR errors cause a single utterance to be transcribed incorrectly, considering the preceding utterances or the situational context can help mitigate or correct such errors.

%% file: sections/03.MutterMeter_Overview.tex
\section{MutterMeter Overview}

\begin{figure}[t]
    \centering
    \includegraphics[width=0.9\textwidth]{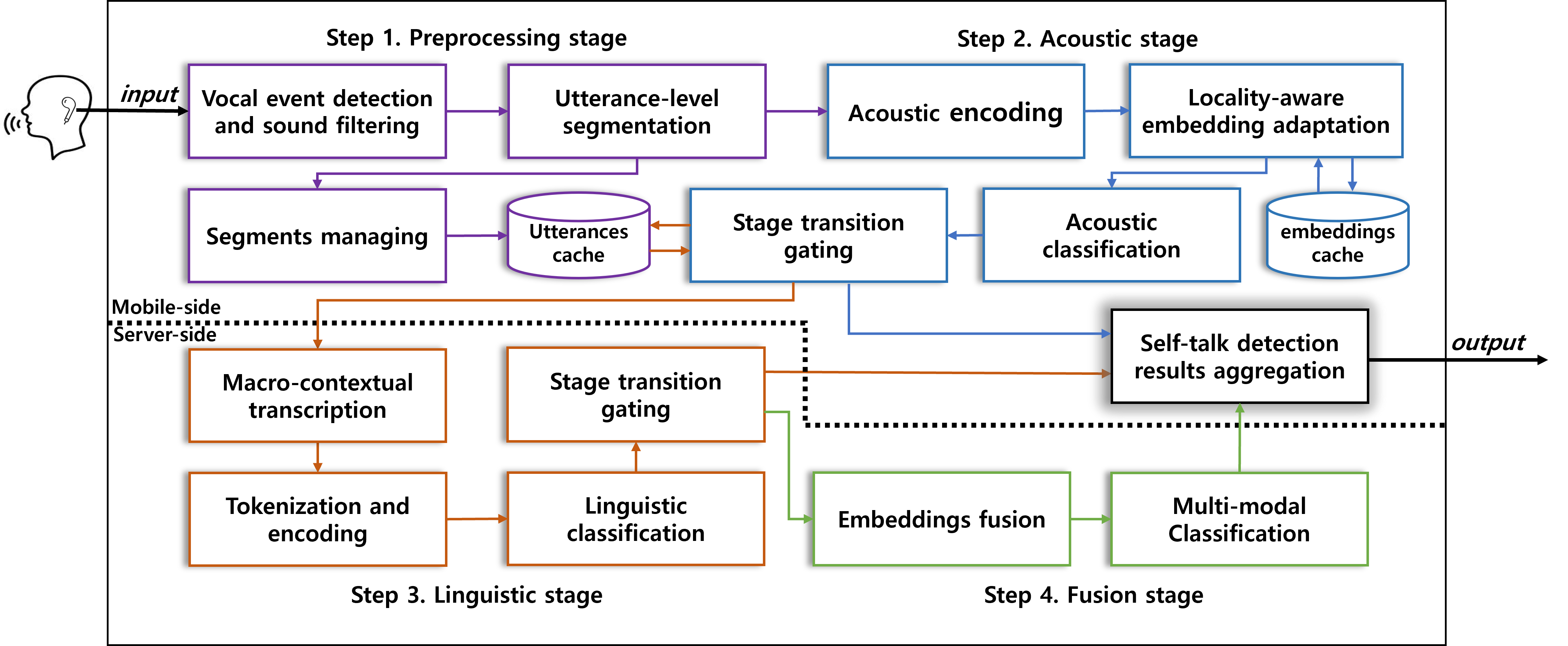}
    \vspace{-0.1in}
    \caption{MutterMeter system overview ~\label{fig:system_architecture}}
    \vspace{-0.15in}
\end{figure}

To effectively detect self-talk from audio signals captured by earables, both acoustic and linguistic features can be useful. However, how to integrate them should be carefully designed. A naive multimodal fusion for all input segments would be computationally inefficient and may even reduce detection accuracy, as the most informative modality differs across segments. To address this, we employ a hierarchical classification-based architecture that dynamically balances accuracy and efficiency while progressively leveraging temporal and contextual information. It employs a sequential processing pipeline consisting of the \textit{acoustic stage}, the \textit{linguistic stage}, and the \textit{fusion stage}, and adaptively decides whether to proceed to the next stage based on classification confidence at each stage (\S\ref{stage_transition_gating}).

Figure~\ref{fig:system_architecture} presents an overview of MutterMeter, which detects self-talk from audio signals captured by user-worn earables. The system consists of a mobile side and a server side, distributed by computational load. The mobile side handles relatively lightweight processing, i.e., receiving audio signals, segmenting them, and performing acoustic classification while exploiting short-term contextual continuity. 
The server side performs computationally intensive tasks, including transcription-based linguistic classification that captures macro-contextual dependencies across utterances, and multimodal fusion that integrates both modalities for final classification.

The main operations of MutterMeter are organized as follows, by stage:
\begin{enumerate}[leftmargin=*]
  \item  \textbf{Preprocessing stage} ($\S\ref{preprocessing_stage}$): This stage precisely segments the input audio into utterance segments and manages them in fixed windows (e.g., 30 seconds) for caching, enabling contextual transcription in a later stage. 
  \item \textbf{Acoustic stage} (\S\ref{acoustic_stage}): This stage generates acoustic embeddings, refines them using locality-aware adaptation to capture short-term contextual continuity, and performs classification. Classification confidence is evaluated to decide whether to move to the next stage.
  \item \textbf{Linguistic stage} (\S\ref{linguistic_stage}): This stage transcribes utterances incorporating macro-contextual dependency across the utterances, extracts the text of the current utterance segment, and performs linguistic classification. Confidence evaluation again determines if fusion is required.  
  \item \textbf{Fusion stage} (\S\ref{fusion_stage}): This stage integrates acoustic and linguistic embeddings for multimodal classification, producing the final classification output.
\end{enumerate}

We believe the proposed hierarchical classification-based architecture provides a flexible and adaptable framework that can be configured to meet the specific requirements of different application scenarios. For instance, when privacy is a primary concern, self-talk detection can be performed locally on the mobile device with a trade-off in classification performance, ensuring that sensitive data remains on the user’s device. Moreover, when responsiveness is critical, particularly for negative self-talk utterances that require instant feedback, the system can be configured to output results directly on the device, enabling real-time responses without server transmission delays. Such flexible architecture allows the system to dynamically balance privacy preservation, latency, and computational efficiency depending on the operational context of applications.

%% file: sections/04.Self-talk_Sensing.tex
\section{Self-talk Sensing}

\subsection{Preprocessing stage}~\label{preprocessing_stage}

To efficiently and accurately detect self-talk, the preprocessing stage must handle incoming audio signals continuously. We design this stage with three main objectives. First, it should be computationally efficient while quickly identifying key signal segments, maximizing recall to capture even short or subtle instances of self-talk. Second, it should detect non-verbal cues—such as sighs, laughter, and murmurs—that provide essential contextual information even when no clear linguistic form is present. 
Third, it should manage the sequential delivery of utterance segments, enabling subsequent stages to leverage macro-contextual dependencies to interpret utterances within their broader context. 
The preprocessing stage is organized into three modules to meet these objectives, as illustrated in Figure~\ref{fig:system_architecture}.

\subsubsection{Vocal event detection and sound filtering}
Because MutterMeter assumes that users wear earable devices close to the mouth, it adopts a simple and efficient decibel (dB)-based method to detect vocal events. The stable distance between the mouth and the device ensures consistent voice capture at a sufficiently high volume~\cite{lee2023groovemeter}. Leveraging this property, the system extracts audio signals above a certain dB threshold as vocal events while filtering out lower-level signals. Empirically, users’ voices recorded via earable microphones typically exhibited an RMS intensity above –20 dB; thus, signals below this level are excluded from detection. This lightweight approach meets key objectives of the preprocessing stage: it efficiently extracts self-talk-related vocal elements (e.g., interjections, emotional expressions), minimizes computational cost for real-time mobile processing, and automatically filters environmental noise and nearby speech that occur at lower amplitudes. Moreover, it helps prevent interference from external voices that could disrupt the contextual continuity of self-talk.
However, it can also capture transient noises, such as impact sounds from user movement or wind noise, which may occasionally exceed the –20 dB threshold and cause false detections; hence, a subsequent segmentation process refines the extracted vocal events into utterance-level units for classification.

\subsubsection{Utterance-level segmentation}
The second step performs utterance-level segmentation on the filtered vocal signals. We assume that no semantic transition occurs within a single utterance, making it the basic unit for self-talk recognition. When multiple meanings are mixed within a segment, semantic ambiguity and contextual overlap can hinder the model's ability to identify self-directed vocalizations. Thus, precise utterance-level segmentation is critical for reliable detection.

\begin{figure}[t]
    \centering
    \includegraphics[width=0.9\textwidth]{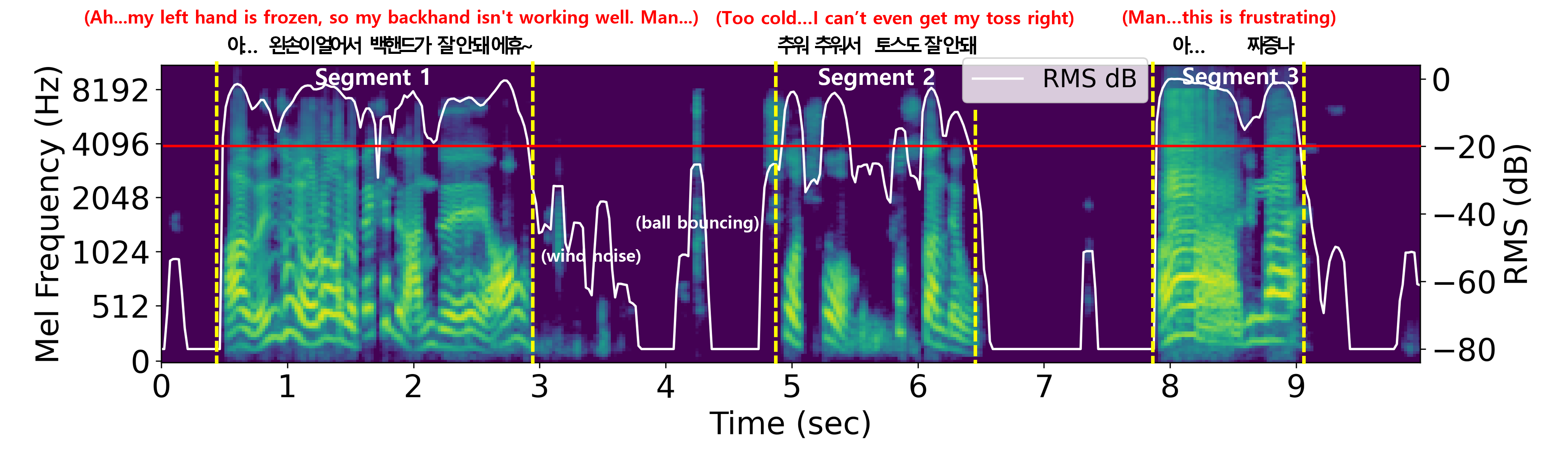}
    \vspace{-0.1in}
    \caption{An example of segmentation result~\label{fig:segmentation}}
    \vspace{-0.2in}
\end{figure}

To achieve accurate utterance-level segmentation, we leverage the temporal characteristics of the filtered audio signals. A simple dB-based threshold often includes meaningless noises or fails to separate utterances properly, requiring a more precise segmentation method. Our method is based on two key observations. First, empirical analysis showed that meaningful vocalizations, including self-talk, always lasted longer than 300 ms; thus, any signal above –20 dB RMS lasting less than 300 ms is discarded as noise. Second, prior linguistic studies~\cite{heldner2010pauses, lee2017speech} reported that the average pause between utterances is about 600 ms in Korean and 500–730 ms in English. Informed by these observations, we define the segmentation rule considering both the signal's duration and interval: multiple segments longer than 300 ms are grouped into a single utterance if their intervals are shorter than 800 ms, whereas longer intervals mark boundaries between utterances. Figure~\ref{fig:segmentation} illustrates this process, where Segments 1–3 correspond to distinct utterances separated by intervals exceeding 800 ms. Notably, Segment 2 would have been incorrectly divided into smaller syllable-level units under simple dB and duration thresholds, but our method correctly combines it into a single meaningful utterance.

\subsubsection{Utterance caching}~\label{utterance_caching}
The segments managing module and utterances cache are designed to provide preceding utterance segments to improve transcription quality at the linguistic stage. This allows the transcription module to utilize not only the current utterance but also macro-contextual information from preceding utterances, enabling the generation of more natural and coherent text. 

The segments managing module dynamically updates the cache whenever a new utterance segment is input, taking into account the total duration of the segments already stored in the cache. Specifically, let the current time step be $t$, the input segment at this time be $s_{t}$, and its duration be $\text{len}(s_{t})$. The set of previously cached segments is denoted as $C_{t-1}$. When a new segment $s_{t}$ is received, the module evaluates the following conditions to determine how to update the cache.
\begin{equation}
\sum_{s_i \in \mathcal{C}_{t-1}} \text{len}(s_i) + \text{len}(s_t) \leq T_{\text{max}}    
\end{equation}
Here, $T_{\max}$ denotes the maximum cacheable duration (e.g., 30 seconds). If the above condition is satisfied, the current segment $s_{t}$ is added to the cache. However, if the total duration of the cached segments including $s_{t}$ exceeds $T_{\max}$, the module sequentially removes the oldest segments until the total duration becomes less than or equal to the maximum duration. Let $C'_{t-1}$ denote the remaining set of segments after this process; then, the cache is updated as follows:
\begin{equation}
\mathcal{C}_{t} = \mathcal{C}_{t-1}' \cup \{s_t\} \quad \text{where} \quad \sum_{s_i \in \mathcal{C}_{t-1}' \cup \{s_t\}} \text{len}(s_i) \leq T_{\text{max}}
\end{equation}
Through this mechanism, the cache is managed to contain as many utterance segments as possible within the maximum duration, currently set to 30 seconds; the rationale behind this decision is described in Section~\ref{linguistic_stage}. The cached segments are transmitted to the server side when self-talk detection for the current utterance is not completed in the acoustic stage.

\subsection{Acoustic stage}~\label{acoustic_stage}

The acoustic stage detects self-talk from utterance-level audio segments using acoustic features through three steps. First, a fine-tuned Transformer-based encoder converts utterances into embeddings for self-talk detection. Second, a locality-aware embedding adaptation method leverages temporal continuity across adjacent utterances to reduce intra-class variation and embedding consistency. Finally, the adapted embeddings are fed into a classification module that determines the utterance class.

\subsubsection{Acoustic encoding}

To generate high-quality acoustic embeddings, we employ the Whisper~\cite{radford2023robust} encoder, a widely used speech encoder. The Whisper encoder is a Transformer-based encoder with a multi-head self-attention architecture, which is advantageous to learn long-range dependencies, allowing it to capture not only contextual information across the entire utterance, but also nonverbal patterns such as intonation and pauses. We fine-tune it on our dataset to produce embeddings appropriate for self-talk detection.

\begin{figure}[t]
    \centering
    \includegraphics[width=0.7\textwidth]{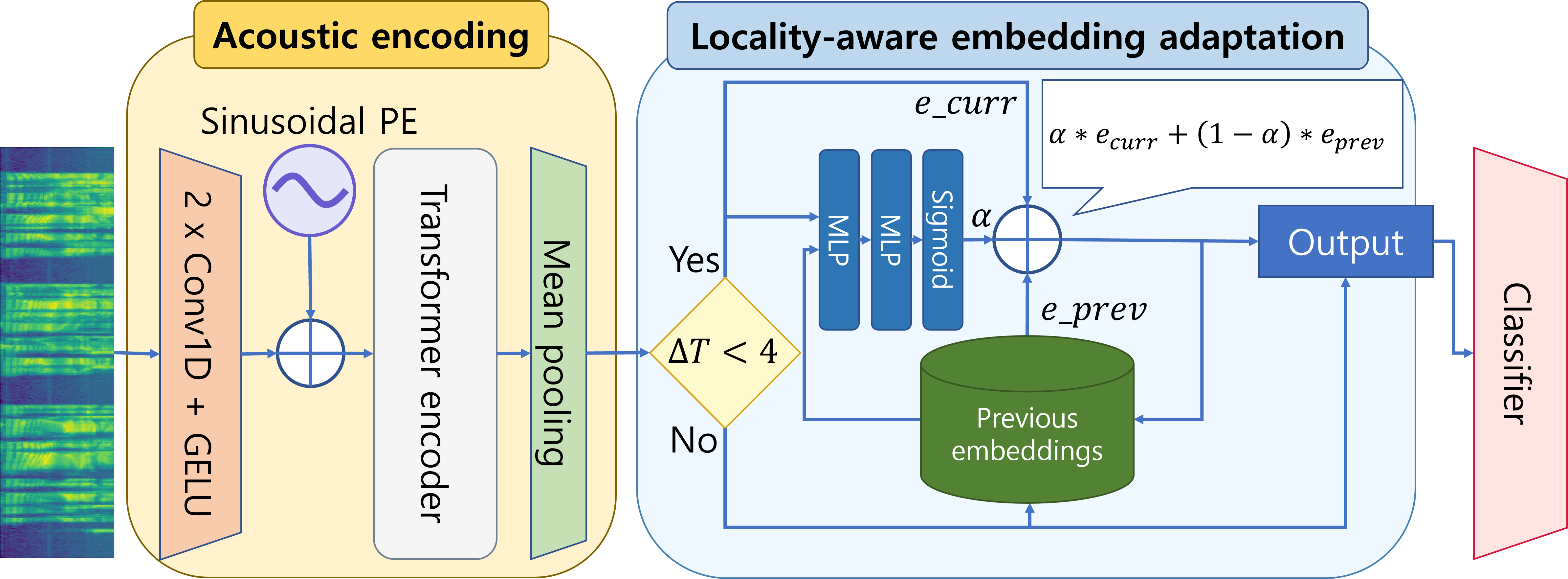}
    \vspace{-0.1in}
    \caption{Acoustic stage self-talk detection pipeline~\label{fig:locality-aware}}
    \vspace{-0.15in}
\end{figure}

The overall encoding process, as illustrated in Figure~\ref{fig:locality-aware}, is structured as follows. First, the raw audio signal segments are converted into 80-channel log-mel spectrograms using the Short-Time Fourier Transform (STFT). The spectrograms are then passed through two Conv1D layers followed by GELU activation to extract low-dimensional local features. To incorporate the sequential order of the input frames, sinusoidal positional encoding is added to these features. Finally, the processed features are fed into the fine-tuned Whisper encoder to generate high-dimensional acoustic embeddings, and mean pooling is applied to produce a single embedding vector for each utterance.

\subsubsection{Locality-aware embedding adaptation}~\label{sec:laea}
To address the challenge posed by the high intra-class variation (see Section~\ref{sec:acoustic_characteristics}), we propose a locality-aware embedding adaptation (LAEA) method that leverages short-term contextual information between utterances (Figure~\ref{fig:locality-aware}). The key idea is that temporally adjacent utterances tend to maintain continuity and are therefore more likely to belong to the same class. By incorporating embeddings of neighboring utterances and applying smoothing, our method reduces intra-class variation and enhances embedding consistency, improving the robustness of self-talk detection.

\begin{figure}[t]
    \centering
    \includegraphics[width=0.8\textwidth]{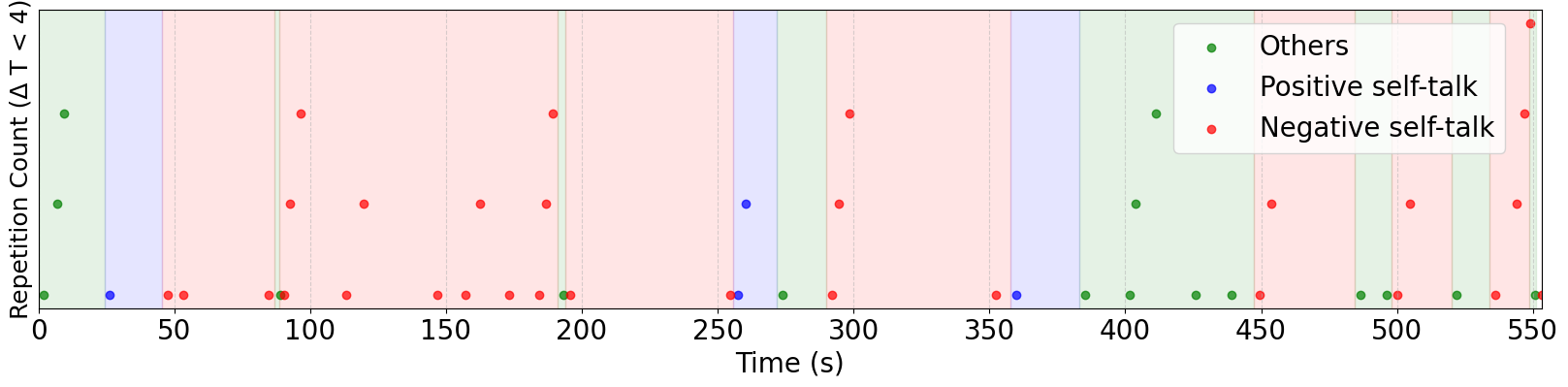}
    \vspace{-0.1in}
    \caption{Repetition patterns of uterances within 4 seconds~\label{fig:repetation_effect}}
    \vspace{-0.15in}
\end{figure}

Our dataset analysis supports this assumption. Approximately 79\% of utterances occurring within four seconds belong to the same class, suggesting strong short-term temporal continuity. Figure~\ref{fig:repetation_effect} illustrates this pattern, where consecutive utterances, particularly negative self-talk, often appear repeatedly within short time intervals. Each point represents an utterance, and the y-axis indicates the number of repetitions of the same class within a four-second window. For example, between 50 and 260 seconds, negative self-talk occurred repeatedly. This empirical finding aligns with the emotional inertia effect observed in psychological studies~\cite{Koval2016}, suggesting that consecutive utterances tend to share the same emotional state.

Motivated by this observation, we design an exponential moving average (EMA)-based embedding adaptation method that leverages consecutive utterances within a 4-second window. Furthermore, we introduce a mechanism that dynamically adjusts the adaptation strength based on the characteristics of each utterance.
Figure~\ref{fig:locality-aware} shows the operation of the EMA-based adaptation. Each utterance embedding is processed in one of two ways, depending on the time interval ($\Delta T$) from the previous utterance. When $\Delta T$ is within 4 seconds, the current utterance embedding ($e_{\text{curr}}$) and the cached embedding from the previous utterance ($e_{\text{prev}}$) are jointly fed into a 2-layer MLP to compute the EMA weight ($\alpha$). The adapted embedding ($e_{\text{adapted}}$) is then generated as a weighted sum of the two embeddings. If $\Delta T$ exceeds 4 seconds, the module skips the adaptation process, as temporal continuity is assumed to be low.
The overall computation is defined as follows:

\begin{equation}
e_{adapted} = \begin{cases}
\alpha \cdot e_{curr} + (1-\alpha) \cdot e_{prev} & \text{if } \Delta T \leq 4 \text{ seconds} \\
e_{curr} & \text{otherwise}
\end{cases}
\end{equation}

\begin{figure}[t]
    \centering

    \begin{subfigure}[b]{0.62\textwidth}
        \centering
        \includegraphics[width=0.48\textwidth]{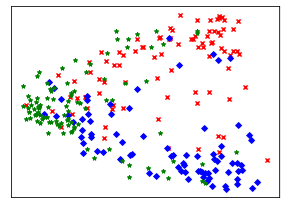}
        \includegraphics[width=0.48\textwidth]{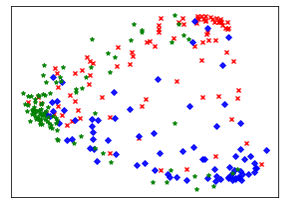}
        \caption{Before and after embedding adaptation for participants P7.}~\label{fig:effect_embedding_adaptation_p7}
    \end{subfigure}
    \hspace{-3 mm}
    \begin{subfigure}[b]{0.38\textwidth}
        \centering
        \includegraphics[width=\textwidth]{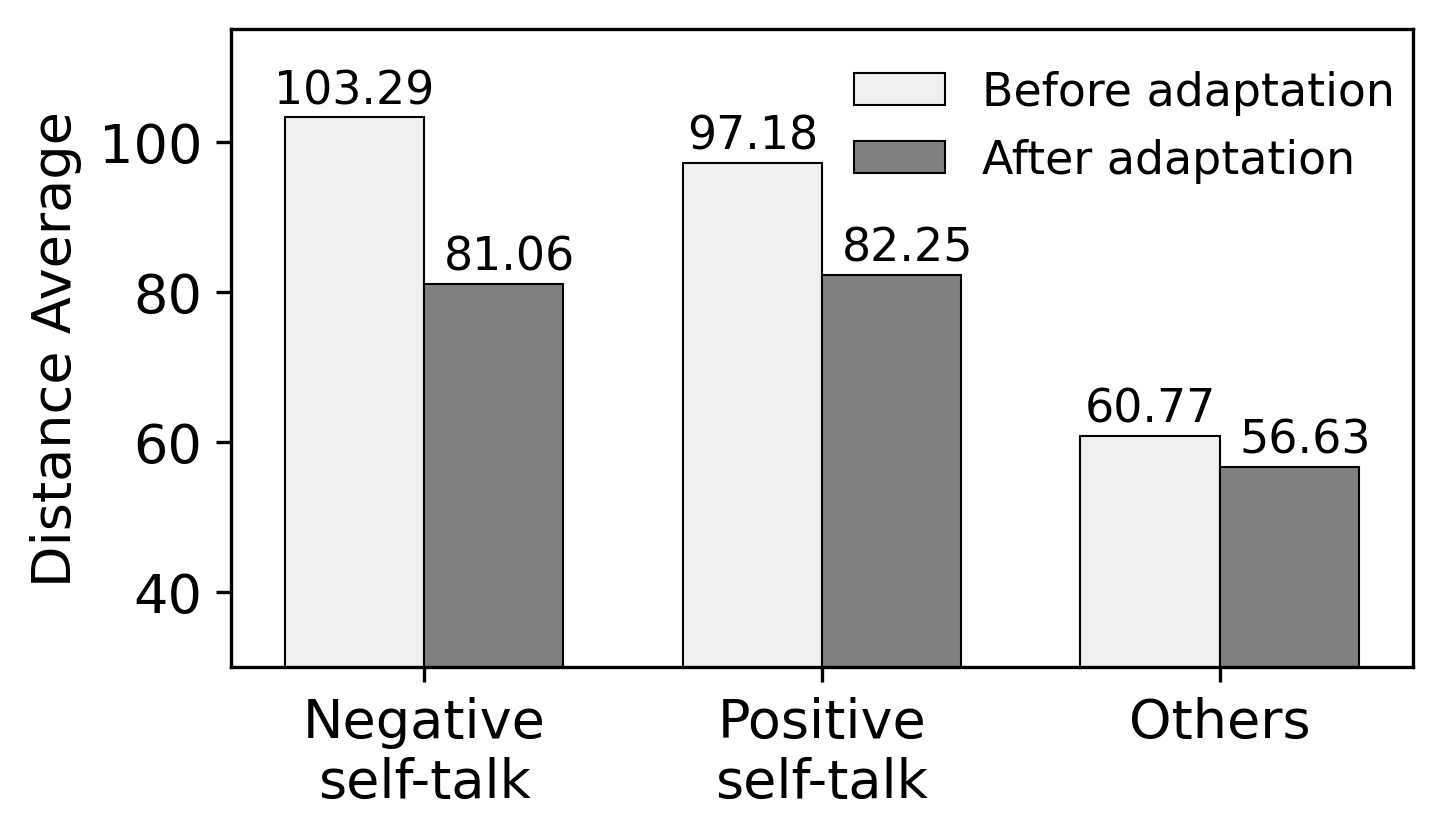}
        \caption{Difference in embedding distances}~\label{fig:average_distance_comparison_pca}
    \end{subfigure}
    \vspace{-0.15in}
    \caption{Effect of LAEA (blue: positive self-talk, red: negative self-talk, green: others).}
    \vspace{-0.15in}
    \label{fig:effect_embedding_adaptation}
\end{figure}

Figure \ref{fig:effect_embedding_adaptation} shows the positive effect of the proposed adaptation method. Figure \ref{fig:effect_embedding_adaptation_p7} presents a PCA visualization of the embeddings for one participant (P7) before and after applying the method. We can observe that embeddings from the same class cluster more tightly following adaptation. Figure \ref{fig:average_distance_comparison_pca} compares the mean distance between embeddings within each class. It shows a considerable reduction in the mean distance after adaptation. 
Notably, the distance reduction is more pronounced for the negative and positive self-talk classes. As presented in Section \ref{sec:background_and_motivation}, self-talk exhibits high intra-class variation, and these results suggest that our embedding adaptation method effectively handles such variability.

\subsubsection{Acoustic classification}
The acoustic classification module takes the adapted embeddings and classifies each utterance into one of three classes. The classifier consists of three fully connected layers with ReLU activation and dropout for regularization, followed by a softmax layer that produces a probability distribution over the classes. The gating module then uses this probability distribution to decide whether to transition to the next stage or finalize the output. Details of the transition gating process are provided in Section~\ref{stage_transition_gating}.

\subsection{Linguistic stage}~\label{linguistic_stage}

The linguistic stage detects self-talk from the transcribed text derived from the input audio. It first generates high-quality transcriptions by leveraging contextual dependency from preceding utterances, then extracts linguistic embeddings using a Transformer-based encoder to capture textual and semantic patterns. Finally, these embeddings are fed into a classification module to determine the utterance type.

\subsubsection{Macro-contextual transcription}

The quality of transcription directly affects the extraction of linguistic embeddings and, consequently, the overall classification performance. A straightforward solution would be to employ a high-performance ASR model, such as Whisper-large with 1.55 billion parameters. However, unsatisfactory transcription quality may still persist, largely due to the inherent characteristics of self-talk.
Unlike typical speech, self-talk often exhibits low volume, unclear articulation, mumbling, and grammatically incomplete sentences (See Section~\ref{sec:acoustic_characteristics}, ~\ref{sec:linguistic_characteristics}). Moreover, our dataset was collected using earable microphones during real tennis play, resulting in audio that contains environmental noise (e.g., wind, clothing friction, and device vibrations) and nonverbal sounds (e.g., breathing, laughter, vocal exertion). These factors collectively degrade transcription quality and hinder the extraction of precise linguistic information.

To improve transcription quality, we leverage macro-contextual dependency surrounding the target utterance. Although recent Transformer-based ASR models (e.g., Whisper) can capture temporal relationships in audio signals via self-attention, their performance can vary depending on input configuration. Since Whisper model has a 30-second input limit, it is crucial to determine how to populate this window with relevant contextual information. To this end, we explore several input composition strategies that organize surrounding segments in different ways, balancing contextual richness against the risk of potential noise contamination.

The most basic strategy directly uses the 30-second audio preceding the target utterance as input. This approach is simple to implement and uses recent contextual information, but it may include irrelevant silence or background noise (\textbf{Contextual transcription w/ prior sound}). A more selective strategy constructs the input using only segments that contain actual utterances within the same session, removing irrelevant segments to emphasize meaningful dependencies between utterances. We devise three variants of the strategy, each balancing temporal proximity and the quantity of utterances included in the input. 
\begin{itemize}
    \item \textbf{Contextual transcription w/o temporal aspects}: The first one fully utilizes the maximum input length regardless of the temporal distance from the current target utterance. Specifically, it aggregates utterance-containing segments from the beginning of the session.  
    \item \textbf{Contextual transcription w/o quantity aspects}: The second one includes only utterances occurring within a short time window (e.g., three minutes) before the target, prioritizing temporal closeness to incorporate contextual information. However, the total input length may not always be fully utilized.
    \item \textbf{Contextual transcription}: The third variant, adopted in MutterMeter, aims to maximize both contextual relevance and input utilization by including as many temporally proximate utterances as possible within the 30-second limit.
\end{itemize}

\begin{table}[t]
\centering
\caption{Transcription quality measured by WER, CER, BLEU, and sacreBLEU.}
\footnotesize
\label{tab:transcription_quality}
\begin{tabular}{l|cccc}
\toprule
& {\textbf{WER $\downarrow$}} & {\textbf{CER $\downarrow$}} & {\textbf{BLEU $\uparrow$}}  & \makecell{\textbf{Sacre} \\ \textbf{BLEU $\uparrow$}}\\
\midrule
 \textbf{Single-utterance transcription} & 0.82 & 1.31 & 0.17 & 23.34 \\
 \textbf{Contextual transcription w/ prior sound} & 0.77 & 1.03 & 0.21 & 23.98 \\
 \textbf{Contextual transcription w/o temporal aspects} & 0.70 & 0.94 & 0.23 & 24.10 \\
 \textbf{Contextual transcription w/o quantity aspects} & 0.69 & 0.88 & 0.27 & 31.47 \\
 \textbf{Contextual transcription} & 0.60 &  0.81 & 0.30 & 36.39 \\

\bottomrule
\end{tabular}
\end{table}

Table~\ref{tab:transcription_quality} summarizes the transcription performance of different input composition strategies, evaluated using WER, CER, BLEU, and sacreBLEU metrics. The results show that contextual transcription consistently achieves the highest performance across all measures. Incorporating contextual information yields better results than transcribing single utterances in isolation. Moreover, utterance-only strategies outperform the contextual transcription w/ prior sound strategy since the 30-second audio window often contained irrelevant silence or background noise that masked useful contextual cues. Contextual transcription w/o temporal aspects enables the model to better capture dependencies among utterances, but sometimes it incorporates irrelevant past utterances, thereby reducing contextual coherence. Contextual transcription w/o quantity aspects emphasizes more recent dependencies, improving local relevance; however, in sparse utterance segments, the absence of preceding utterances within 3 minutes led to insufficient contextual input. Intriguingly, according to BLEU score interpretation guidelines\footnote{\url{https://cloud.google.com/translate/automl/docs/evaluate}}, single-utterance transcription achieves BLEU scores in the range of 0.1–0.2 (“hard to get the gist”), whereas contextual transcription reaches 0.3–0.4 (“understandable to good translations”), confirming its superior transcription quality.

\begin{table}[t]
\centering
\caption{Effect of context-aware transcription; The symbol [ ] denotes a single segment extracted through our segmentation stage, representing one utterance.} 
\label{tab:context_aware_transcription}
\vspace{-0.1in}
\footnotesize
\begin{tabular}{>{\centering\arraybackslash}m{0.5cm} >{\raggedright\arraybackslash}p{3cm} p{11.5cm}}
\toprule
 & \textbf{Method} & \textbf{Transcription results} \\
\midrule
 & \textbf{Ground truth} & [손목을 써야 돼 손목을], [손목을 이렇게 촥! 채는 게 있어야 돼 나도 그게 안 돼 사실], [러브 피프틴!], [댐프너 날아갔다], [댐프너 날아갔어] \newline  \textcolor{purple}{\textit{[I need to use my wrist, use my wrist], [I have to swing my wrist like this properly, but I actually can’t do it], [love, fifteen!], [The dampener flew off], [The dampener flew off] }}\\
\multirow{2}{*}{1} & \textbf{Single-utterance trascription} & [손목을 써야 돼 손목을], [으 으 포모 이렇게 착 채는 게 있어야 돼 나도 그게 안돼 사실], [러피!], [템포도 날아갔다], [램프나 나라가]\newline \textcolor{purple}{\textit{[I need to use my wrist, use my wrist], [Uh uh, I have to swing like this properly, but I actually can’t do it], [Loopy!], [The tempo also flew off], [the Lamp or country] }}\\
& \textbf{Macro-contextual transcription} & [손목을 써야 돼 손목을], [손목을 이렇게 촥! 채는 게 있어야 돼 나도 그게 안 돼 사실], [러브 피프틴!], [댐프너 날아갔다], [댐프너 날아갔다] \newline  \textcolor{purple}{\textit{[I need to use my wrist, use my wrist], [I have to swing my wrist like this, but I can’t actually do it], [Love-fifteen!], [The dampener flew off], [The dampener flew off]}} \\
\midrule
 & \textbf{Ground truth} & [아 이거 뭐야..], [아 씨], [얘 왜 길까]\newline  \textcolor{purple}{\textit{[Ah, what is this…], [Ah, damn], [why is this one so long?]}}\\
\multirow{2}{*}{2} &  \textbf{Single-utterance transcription} & [ああ], [아…], [얘 왜 길까]\newline  \textcolor{purple}{\textit{[ - ], [Ah...], [Why is this one so long?]}}\\
 & \textbf{Macro-contextual transcription} & [아 이거 뭔..], [아 씨], [얘 왜 길까]\newline \textcolor{purple}{\textit{[Ah, what is this…], [Ah, damn], [why is this one so long?]}} \\
\bottomrule
\end{tabular}
\end{table}

Table~\ref{tab:context_aware_transcription} demonstrates that applying contextual transcription produces higher-quality transcription compared to using only a single utterance segment as input. For example, in example 1, the single-utterance transcription produced “Uh uh, I have to swing like this properly, but I actually can’t do it.” instead of the correct “I have to swing my wrist like this properly, but I actually can’t do it.” In this case, the target of “swing”—“wrist”—was omitted, resulting in an ambiguous meaning, and an unnecessary interjection, “Uh uh,” was incorrectly inserted. Furthermore, domain-specific expressions were often misrecognized as entirely different words with similar pronunciations. For instance, “Love-fifteen,” which denotes a score in tennis, was misrecognized as “Loopy!”, while “Dampener,” referring to a vibration-reducing device for a racket, was transcribed as “Tempo” or “Lamp.”

\subsubsection{Tokenization/encoding and classification}

In the tokenization and encoding module, linguistic features are extracted from the transcribed text. Specifically, the part corresponding to the current utterance is cropped from the transcribed text and converted into a linguistic embedding using a BERT-based tokenizer and encoder. The [CLS] token vector is then used as the classifier's input.

In the linguistic classification module, self-talk detection is performed based on the input [CLS] token vector. The classifier consists of three fully connected layers with ReLU activation functions and dropout. Finally, a softmax function is used to compute the probability distribution over three classes. The output is passed to the stage transition gating module, where the result from the linguistic stage is either finalized as the output or passed to the fusion stage.

\subsection{Fusion stage}~\label{fusion_stage}

The fusion stage, the final component of the MutterMeter architecture, is designed to further enhance the accuracy of self-talk classification. This stage handles segments that are difficult to classify reliably using a single modality, as each modality captures only partial information about the signal. To address this limitation, the fusion stage receives embeddings extracted from both modalities in the previous stages. It then integrates them to address incomplete representations from individual modalities, thereby improving overall classification reliability.

For effective modality fusion, the fusion stage is designed to adaptively determine the weighting of each modality based on the characteristics of each segment. As discussed in Section~\ref{sec:preliminary_exploration}, acoustic features play a more decisive role in some cases, whereas linguistic features are more informative in others. Moreover, when transcriptions become unreliable due to noise or unclear articulation, linguistic information may become less effective. Therefore, applying a fixed weighting scheme to both modalities has inherent limitations; instead, an adaptive fusion mechanism that dynamically adjusts modality importance per segment is essential for achieving robust performance.

To realize this adaptive mechanism, we adopt a gated fusion approach~\cite{arevalo2017gated, li2020gated}. Specifically, before fusing modality-specific features for each segment, the fusion stage applies a gating mechanism to implement adaptive weighting, adjusting each modality’s contribution according to segment characteristics. For instance, when the gate value is close to 1, the information from the corresponding modality is almost fully reflected, whereas when it is close to 0, that information is almost entirely ignored. The sum of the gate values for the two modalities is constrained to 1. Values between 0 and 1 allow partial information to pass through, enabling the fusion stage to effectively capture the varying importance of each modality across different segments.
Specifically, given acoustic embedding $x_1 \in E_a$ and linguistic embedding $x_2 \in E_l$, both projected to the same dimensionality, the gating weight $g$ is computed as:
\begin{equation}
g = \sigma(W_g [x_1; x_2] + b_g)
\end{equation}
Here, $\sigma$ denotes the sigmoid function, and $[x_1; x_2]$ represents the concatenation of the two embeddings. $W_g$ and $b_g$ refer to the learnable weight matrix and bias term for the gating operation, respectively. The final fused feature $z$ is then obtained as:
\begin{equation}
z = g \odot x_1 + (1 - g) \odot x_2
\end{equation}
In this formulation, $\odot$ indicates element-wise multiplication, allowing the model to adaptively control the contribution of each modality for every segment. The fused feature $z$ is subsequently fed into a classifier consisting of five fully connected layers with ReLU activations and Dropout applied between each layer, to classify into the three target classes.

\subsection{Stage transition gating}~\label{stage_transition_gating}

MutterMeter adopts a hierarchical classification approach to balance accuracy and computational efficiency in self-talk detection. Instead of jointly processing acoustic and linguistic modalities at a single stage, each segment is progressively processed through the acoustic, linguistic, and fusion stages, advancing only when additional information is required. Such an approach allows the system to adaptively allocate computational resources to only the segments that require additional processing, thereby ensuring efficient yet reliable classification performance across diverse self-talk segments.

To implement the hierarchical approach, MutterMeter employs a confidence-guided stage transition gating mechanism. Each segment first undergoes classification in the acoustic stage, where predictions are made solely from acoustic features. If the model’s confidence in a prediction exceeds a predefined threshold, the result is finalized; otherwise, the segment proceeds to the linguistic stage, where linguistic information refines the decision. When uncertainty persists even after linguistic processing, the fusion stage combines both modalities to produce a final prediction. For confidence-guided gating, we use the least margin—the difference between the two highest class probabilities—as the confidence score, which quantifies the model’s certainty in its prediction. It is important to carefully determine confidence threshold values tuned to balance accuracy and efficiency. Improper threshold settings can cause premature exits from early stages, increasing the likelihood of errors when the early decisions are wrong, or trigger unnecessary transitions to subsequent stages, leading to excessive computational costs such as unnecessary transcription operations.

\begin{figure}[t]
    \centering

    \subfloat[Acoustic stage\label{fig:least_margin_audio}]{
        \includegraphics[width=0.5\textwidth]{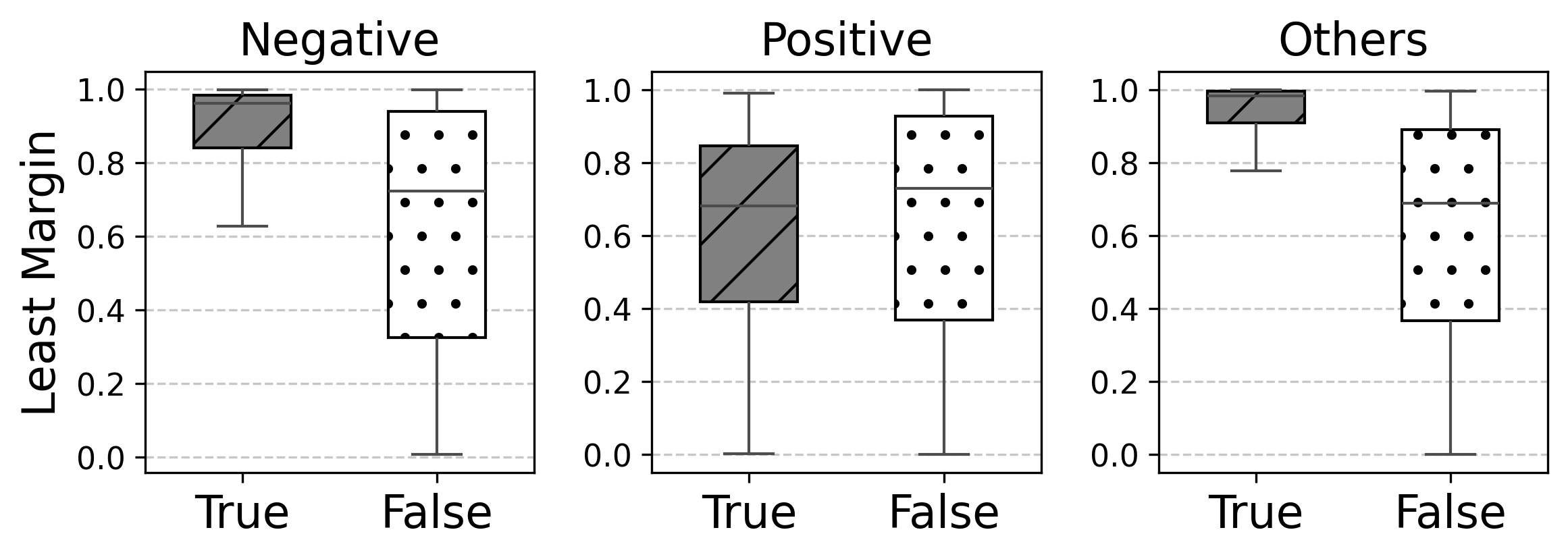}
    }
    \subfloat[Linguistic stage\label{fig:least_margin_text}]{
        \includegraphics[width=0.5\textwidth]{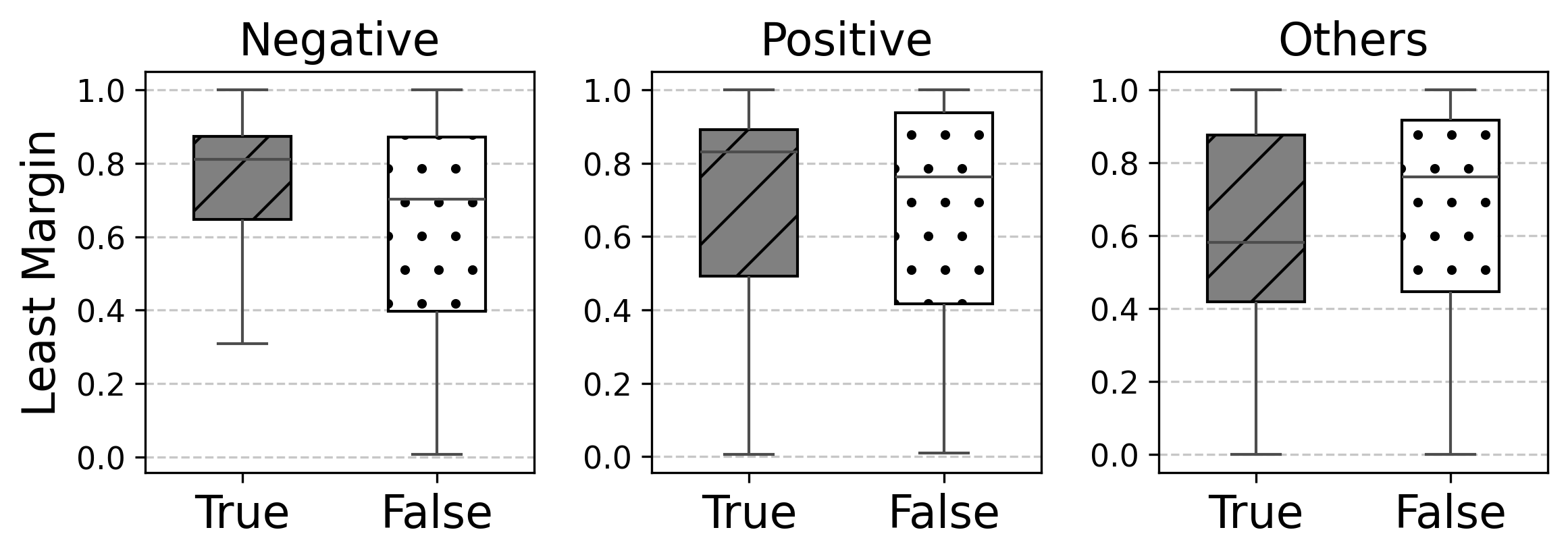}
    }
    \vspace{-0.15in}
    \caption{Least margin distribution\label{fig:least_margin}}
\end{figure}

\begin{algorithm}[t]
\footnotesize
\caption{Stage transition gating}
\label{algo:transition_gating}

\SetKwInOut{Input}{Input}
\SetKwInOut{Output}{Output}

\Input{Input segment $x$}
\Output{Final prediction $\hat{y}$}

\BlankLine
\textbf{Notation:} \;
$y_s$: prediction result at stage $s$ \;
$m_s$: least margin at stage $s$ \;
\BlankLine

$y_\text{Acoustic}, m_\text{Acoustic} \gets \text{Acoustic stage}(x)$ \;
\If{$(y_\text{Acoustic} \in \{\text{negative, other}\})$ \textbf{and} $(m_\text{Acoustic} \ge 0.92)$}{
    $\hat{y} \gets y_\text{Acoustic}$ \;
}
\Else{
    $y_\text{Linguistic}, m_\text{Linguistic} \gets \text{Linguistic stage}(x)$ \;
    \If{$(y_\text{Linguistic} = \text{negative})$ \textbf{and} $(m_\text{Linguistic} \ge 0.80)$}{
        $\hat{y} \gets y_\text{Linguistic}$ \;
    }
    \Else{
        $y_\text{Fusion} \gets \text{Fusion stage}(x)$ \;
        $\hat{y} \gets y_\text{Fusion}$ \;
    }
}

\Return $\hat{y}$ \;
\end{algorithm}

We employ adaptive thresholds determined based on the empirical distributions of class-wise confidence scores. Figure~\ref{fig:least_margin} illustrates the class-wise distribution of least margins in the acoustic and linguistic stages, showing distinct patterns across both stages and classes, which are obtained from our dataset. Based on these observations, we define the transition conditions for each stage as presented in Algorithm~\ref{algo:transition_gating}. Specifically, in the acoustic stage, if the prediction result is negative or other and the least margin is ≥ 0.92, the prediction is accepted as the final output; otherwise, the process transitions to the linguistic stage. In this stage, if the prediction is negative and the least margin is ≥ 0.80, the result is accepted as final; otherwise, it proceeds to the next stage. The fusion stage, as the final layer, processes all remaining segments and produces the final classification results.

%% file: sections/05.DataCollection.tex
\section{Data Collection}~\label{sec:data_collection}

To build and evaluate MutterMeter, we collected a new dataset comprising various speech utterances, including self-talk, recorded during tennis play. To the best of our knowledge, this is the first dataset specifically targeting self-talk.
The entire data collection process was carried out under the approval of the Institutional Review Board.

\subsection{Participants}
We recruited 25 participants (22 male, 3 female) who regularly play tennis, including members of a university tennis club, a local community tennis club, and high school tennis players. We directly contacted the organizers of the university and community clubs, as well as a high school tennis coach. We asked them to carefully recommend potential participants for the data collection, taking into account their skill levels, physical condition, and overall suitability. Note that the participants had an average of 2.4 years of playing experience and played tennis approximately 2.7 times per week. Each participant received a base compensation of approximately USD 8 per hour. To further motivate participants and encourage active engagement during matches, we informed them in advance that an additional reward of approximately USD 4 would be awarded to the winner of each set, which was provided accordingly.

\begin{figure}[t]
    \centering

    \subfloat[Audio recording\label{fig:earphone}]{
        \includegraphics[width=0.35\textwidth]{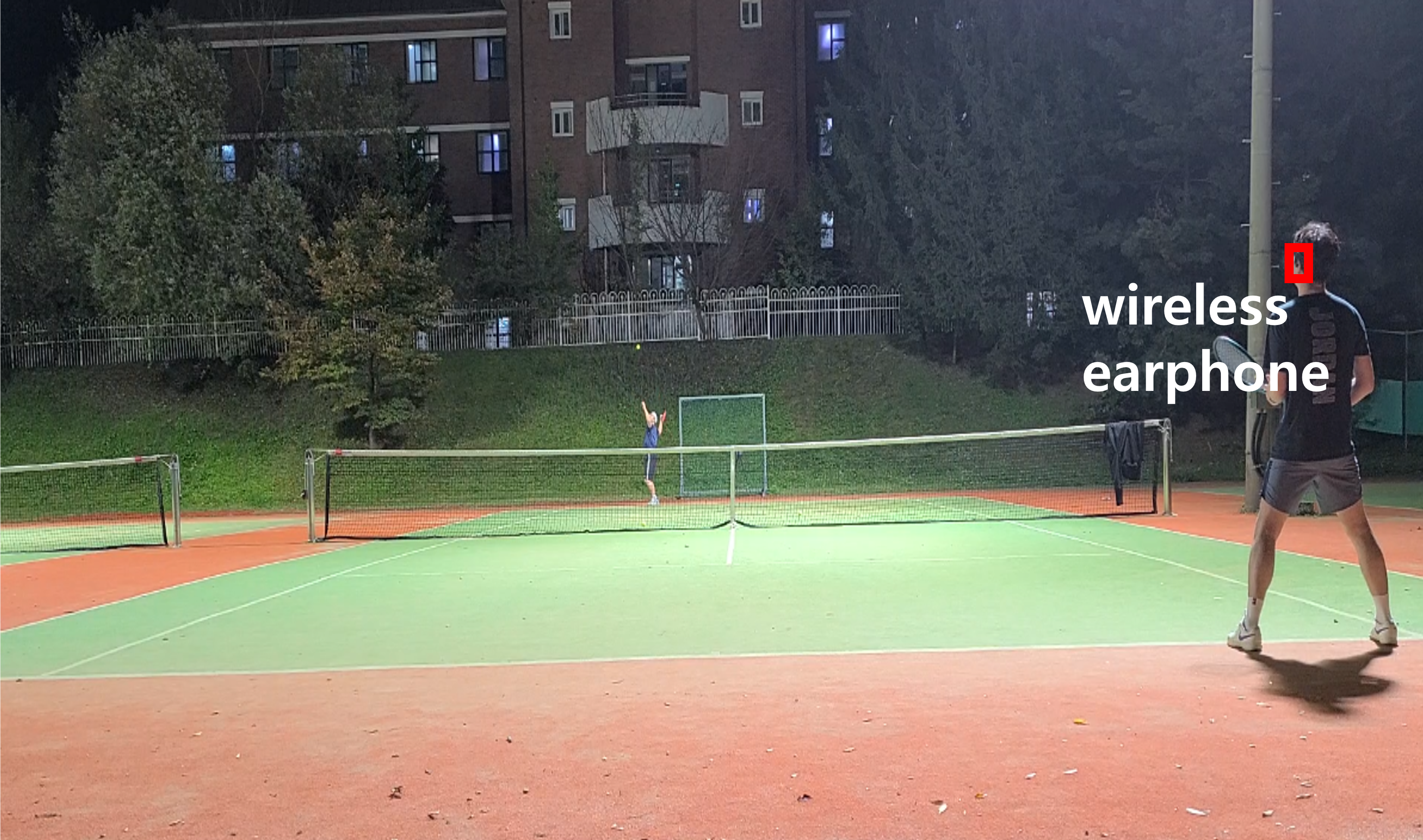}
    }
    \hspace{0.05\textwidth}
    \subfloat[Video recording\label{fig:camera_location}]{
        \includegraphics[width=0.35\textwidth]{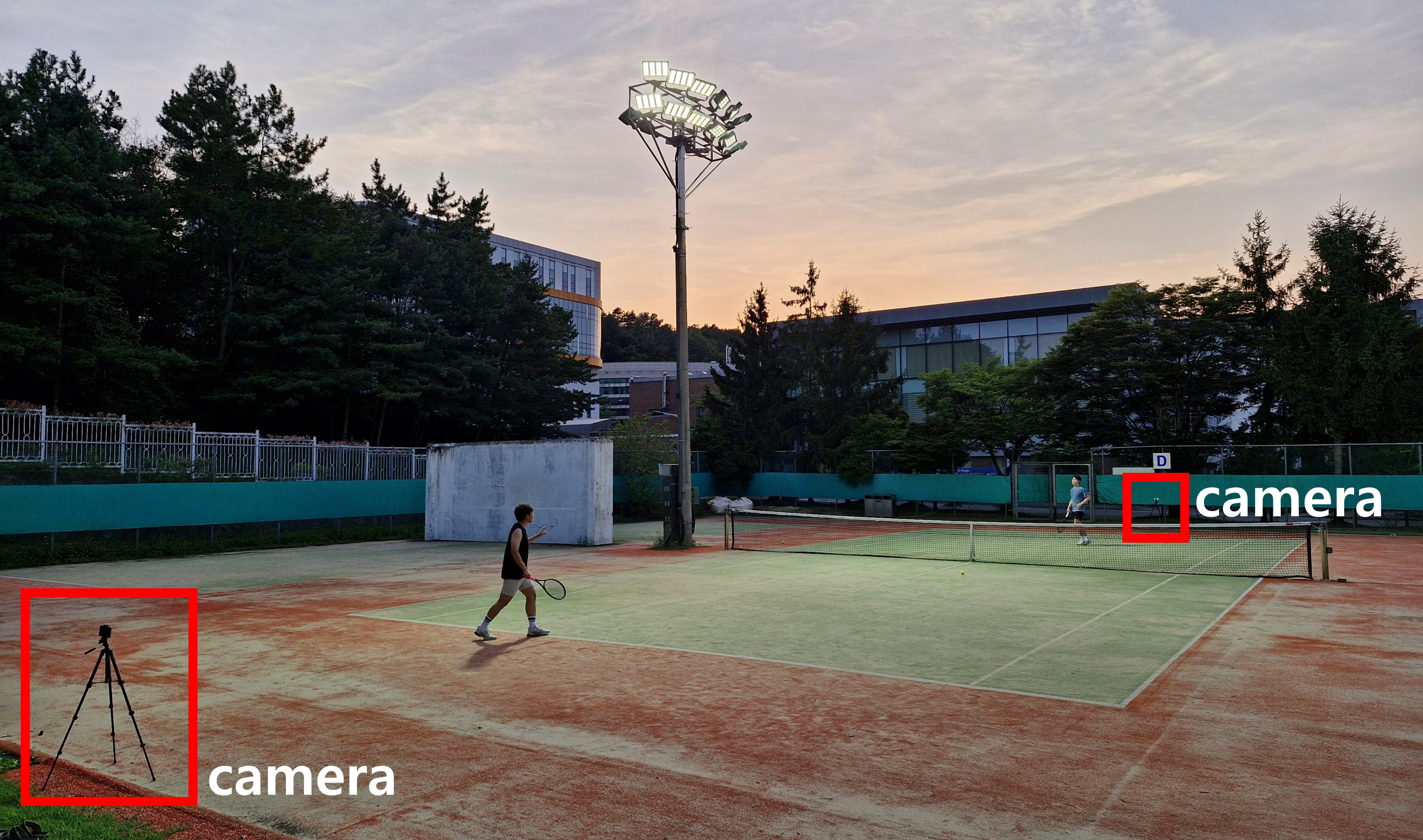}
    }
    \vspace{-0.1in}
    \caption{In-the-wild data collection setup\label{fig:camera_gt}}
    \vspace{-0.15in}
\end{figure}

\subsection{In-the-wild data collection}
Each participant was invited to a tennis court arranged by the researchers. Note that all courts used for the data collection were public facilities located within universities or local communities. Upon arrival, participants were provided with data collection equipment (e.g., wireless earphones, cameras) and the custom mobile application, followed by a detailed explanation of the experimental procedures. We then obtained informed consent from the participants. Afterward, participants engaged in a brief warm-up rally to relieve tension and to familiarize themselves with the equipment and the app.

During the main data collection sessions, the researchers minimized intervention unless technical issues with the equipment or the app occurred. Participants freely played three full sets of tennis matches. The use of wireless earphones for recording and video cameras continued throughout the entire session, from the beginning to the end of each match. We were able to naturally capture not only all on-court moments during play but also various ambient sounds and interactions during breaks, such as conversations between participants, chatting with spectators, and the sounds of drinking beverages, thereby reflecting the natural acoustic environment of real tennis.

Additionally, we did not control the surrounding environment during data collection. For example, other individuals unrelated to the data collection were free to play tennis on adjacent courts while the data collection was ongoing. The recordings, therefore, naturally contained unfiltered ambient sounds, including their conversations, cheers, and ball-hitting sounds. In some cases, when data were collected on university courts, announcements from the campus broadcasting system were also captured. Similarly, for courts located near busy roads, ambient noise such as motorcycle engines and police sirens was recorded.

To collect audio data, we used wireless earphones connected to an Android phone (see Figure~\ref{fig:earphone}). Specifically, we employed open-type wireless earphones (Britz BZ-D100X\footnote{\url{https://britz.co.kr/product/view.html?from=all&p_no=1113}}
) to allow participants to hear external sounds without difficulty and minimize interference with tennis performance. Note that, according to a survey conducted immediately after the data collection, participants reported minimal discomfort while wearing the earphones during tennis matches, with an average score of 5.8 (SD = 1.6) on a 7-point scale. The sampling rate of the earphone microphone was set to 22,050 kHz.
Additionally, to obtain ground truth and capture the overall context of speech events, we installed one camera per participant to record their actions during tennis matches. The cameras were positioned as shown in Figure~\ref{fig:camera_location} to ensure minimal interference with gameplay.

\subsection{Ground truth labeling and summary statistics}

We conducted manual labeling and transcription of the collected dataset to construct ground truth annotations, i.e., triplets of <timestamp, utterance text, utterance type>. Labeling was performed by listening to the audio recordings and identifying auditory and semantic cues within each utterance. When the annotation was ambiguous, the video recordings were reviewed to interpret the situational context. Furthermore, to ensure reliability, two researchers collaboratively tagged the ground truth for all segments, discussing segments as needed to finalize the labels. 
The labeling process was facilitated using a custom-built tagging tool that synchronized audio–video playback and labeling panels for efficient annotation.

\begin{figure}[t]
  \centering

  \begin{subfigure}[b]{0.25\textwidth}
    \includegraphics[width=\linewidth]{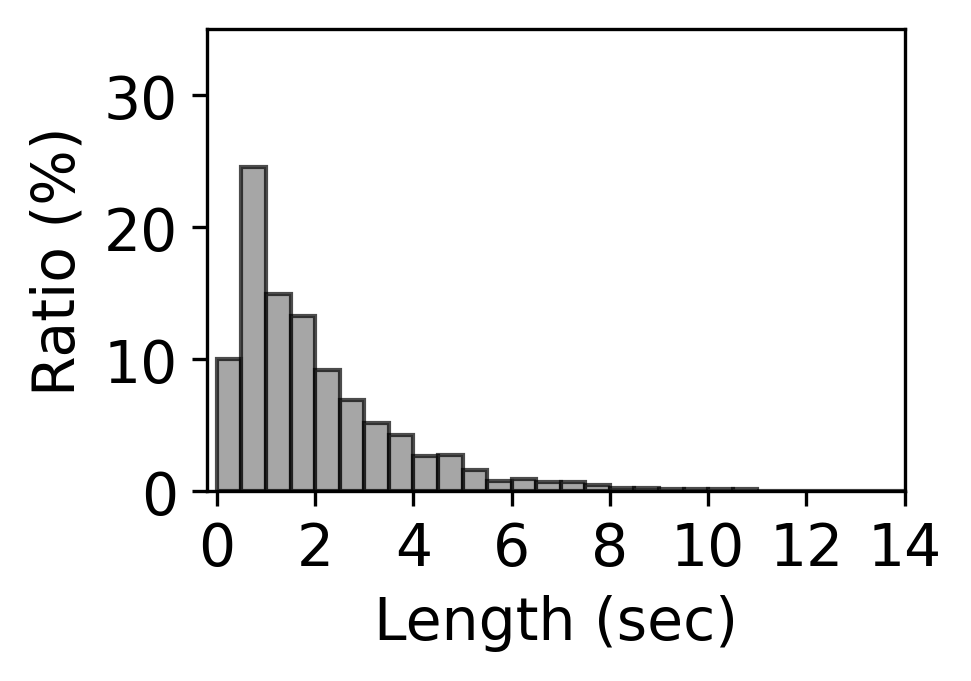}
    \caption{Negative self-talk}
    \label{fig:negative_histogram}
  \end{subfigure}
  \begin{subfigure}[b]{0.25\textwidth}
    \includegraphics[width=\linewidth]{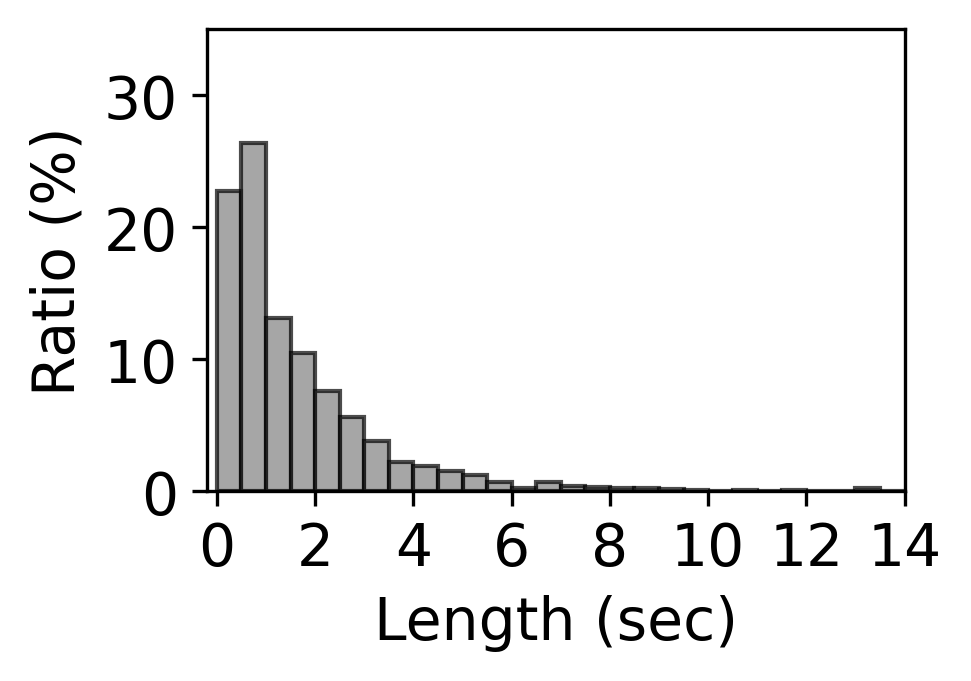}
    \caption{Positive self-talk}
    \label{fig:positive_histogram}
  \end{subfigure}
  \begin{subfigure}[b]{0.25\textwidth}
    \includegraphics[width=\linewidth]{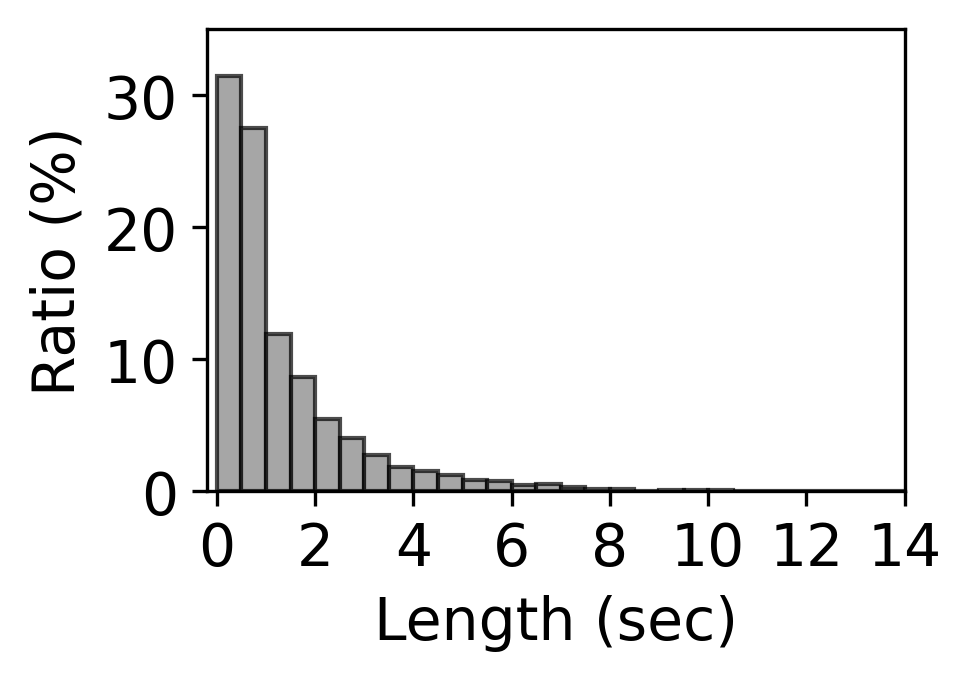}
    \caption{Non self-talk}
    \label{fig:others_histogram}
  \end{subfigure}

  \vspace{-0.1in}
  \caption{Distribution of utterance durations}  
  \label{fig:duration_histogram}
    \vspace{-0.1in}
\end{figure}

We collected approximately 31.1 hours of data from 25 participants, including 8,900 utterance segments with a cumulative duration of approximately 3.9 hours. Among these segments, 26\%, 16\%, and 58\% corresponded to negative self-talk, positive self-talk, and others, respectively. Figure~\ref{fig:duration_histogram} shows the histograms of duration for the three utterance types, indicating how long each utterance session lasted. The mean duration of each utterance was 2.0 s (SD = 1.8, MAX = 13.9, MIN = 0.3) for negative self-talk, 1.6 s (SD = 1.7, MAX = 19.9, MIN = 0.3) for positive self-talk, and 1.4 s (SD = 1.5, MAX = 16.0, MIN = 0.3) for other utterances. Negative self-talk generally exhibits longer utterances, as participants often engaged in pessimistic self-talk when losing or facing challenging moments in the game. Most utterances are shorter than 2 seconds. In particular, over 50\% of non-self-talk utterances lasted less than 1 second, reflecting frequent short communications such as score announcements or in/out calls during tennis play.

%% file: sections/06.Evaluation.tex
\section{Evaluation}~\label{sec:evaluation}

\subsection{Model training and testing}

The collected dataset was processed following the procedure described in the preprocessing stage (Section~\ref{preprocessing_stage}), and the resulting utterance-level segmented audio signals were used for both training and testing. To evaluate the performance of self-talk detection, we conducted a leave-one-subject-out (LOSO) cross-validation (CV) on 25 participants. All training and testing was conducted using the PyTorch framework on an NVIDIA TITAN RTX environment.

In the acoustic stage, the acoustic encoder was fine-tuned based on the Whisper-base~\footnote{\url{https://huggingface.co/openai/whisper-base}} encoder. During training, the batch size was set to 32, the optimizer was AdamW, and the learning rate was $2 \times 10^{-5}$ with $\beta_1 = 0.9$, $\beta_2 = 0.999$, and $\epsilon = 10^{-0.8}$. To prevent overfitting, an early stopping strategy was applied. The 2-layer MLP for the locality-aware embedding adaptation was trained with a batch size of 64, using the Adam optimizer and a learning rate of $1 \times 10^{-4}$. Note that the encoder and locality-aware embedding adaptation model trained with PyTorch were converted to TensorFlow Lite to enable system cost analysis.

In the linguistic stage, the segmented audio signals were transcribed using Whisper-large-v3~\footnote{\url{https://huggingface.co/openai/whisper-large-v3}} and utilized for both training and testing. Since most of the collected data consisted of Korean utterances, we fine-tuned a Korean BERT-base model~\footnote{\url{https://huggingface.co/beomi/kcbert-base}} for tokenization and encoding. During training, the batch size was set to 64, the optimizer was AdamW, and the learning rate was $2 \times 10^{-5}$ with $\beta_1 = 0.9$, $\beta_2 = 0.999$, $\epsilon = 10^{-0.8}$, and weight decay = $0.01$. To prevent overfitting, an early stopping strategy was applied. 

In the fusion stage, embeddings from the acoustic and linguistic stages were used as input, with a batch size of 64. The Adam optimizer was employed with a learning rate of $1 \times 10^{-4}$. To prevent overfitting, an early stopping strategy was applied.

\subsection{Overall detection performance}

\begin{figure}[t]
    \hspace{-0.85in}
    \begin{minipage}[c]{0.4\textwidth}
        \raggedright
        \includegraphics[width=\textwidth]{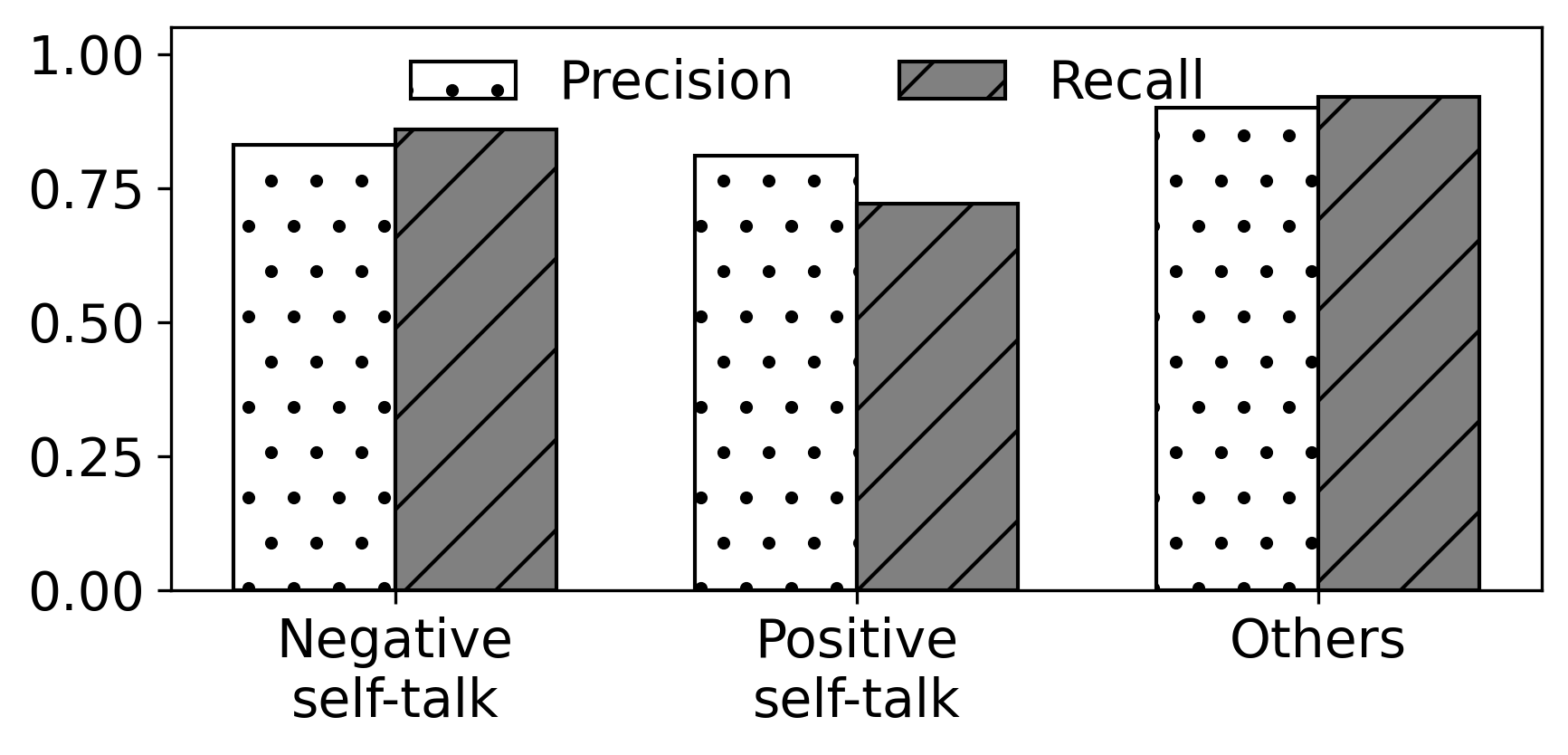}
        \caption{MutterMeter overall performance~\label{fig:pre_rec_overall}}
    \end{minipage}
    \begin{minipage}[c]{0.5\textwidth}

        \captionof{table}{Overall self-talk detection performance ($F_1$ score)}~\label{tab:overall_performance}
        \footnotesize
        \begin{tabular}{ c c c |c c c | c}
        \toprule
        \makecell{\textbf{Acoustic} \\ \textbf{model}} & \makecell{\textbf{Linguistic} \\ \textbf{model}} & \makecell{\textbf{Fusion} \\ \textbf{model}} &  \makecell{\textbf{Negative} \\ \textbf{self-talk}} & \makecell{\textbf{Positive} \\ \textbf{self-talk}} & \textbf{Others}  & \makecell{\textbf{Macro} \\ \textbf{Avg.}} \\
        \midrule
        
        $\bigcirc$ & - & - & 0.81 & 0.71 & 0.89 & 0.80 \\
        - & $\bigcirc$ & - & 0.71 & 0.58 & 0.81 & 0.70 \\
        - & $ - $ & $\bigcirc$ & 0.81 & 0.75 & 0.89 & 0.82 \\
        \rowcolor{gray!20} 
        $\bigcirc$ & $\bigcirc$ & $\bigcirc$ & 0.84 & 0.77 & 0.91 & 0.84 \\
        
        \bottomrule
        \end{tabular}
        
        \label{tab:f1score}
    \end{minipage}
\end{figure}

\begin{figure}[t]
    \centering
    \includegraphics[width=0.80\textwidth]{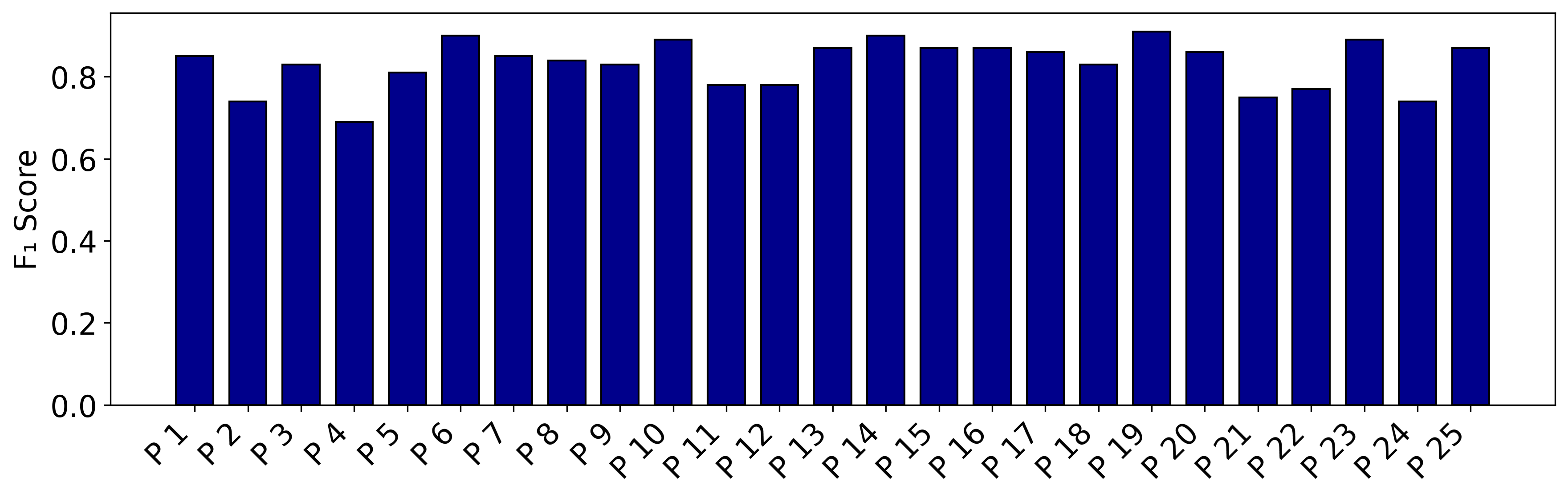}
    \vspace{-0.1in}
    \caption{Performance per participant~\label{fig:per_participant}}
    \vspace{-0.15in}
\end{figure}

Figure~\ref{fig:pre_rec_overall} shows that MutterMeter fairly accurately detects self-talks for unseen users from the in-the-wild audio dataset, achieving the macro-averaged $F_1$ score of 0.84.
More specifically, MutterMeter achieves a precision of 0.83 and a recall of 0.86 for negative self-talk, demonstrating stable detection performance. Positive self-talk shows slightly lower recall, mainly due to linguistic and acoustic similarities with conversational speech. Positive self-talk often manifests as command-like expressions (e.g., “Do it now,” “Focus more”) that resemble instructions to directed toward another person. It also frequently omits explicit subjects, which increases ambiguity and makes interpretation difficult even for human annotators. Furthermore, since positive self-talk often occurs in isolation, the advantage of the locality-aware embedding adaptation becomes limited.
For others, MutterMeter achieves 0.90 precision and 0.92 recall, demonstrating its effectiveness in accurately detecting non–self-talk

MutterMeter generally achieves stable detection performance across participants, as shown in Figure~\ref{fig:per_participant}.
For the participant with the lowest performance (P4), the macro-averaged $F_1$ score is 0.69, while the participant with the highest performance (P19) achieves 0.90.
Participants with lower performance generally show reduced accuracy in detecting positive self-talk. Our observations indicate that some participants (P4, P24) frequently used distinctive personal utterances or vocal habits as forms of positive self-talk, which likely made it difficult to distinguish them from others.
In addition, because the number of female participants is relatively small, some (P21, P22) tend to show lower performance.
Meanwhile, the number of utterances per participant did not have a significant effect on detection performance.

Table~\ref{tab:overall_performance} shows that the proposed MutterMeter's hierarchical self-talk detection model achieves the highest performance across all classes, compared to single-modality models (acoustic, linguistic) and the fusion model.
Specifically, the acoustic model shows inferior performance in detecting positive self-talk, achieving an $F_1$ score of 0.71 compared to 0.77 with the hierarchical approach, an improvement of approximately 8.5\%.
The linguistic model exhibits generally lower performance across all classes, with a macro-averaged $F_1$ score of 0.70, whereas MutterMeter achieves 0.84, representing an improvement of about 20\%.
Interestingly, MutterMeter outperforms fusion models that integrate acoustic and linguistic features for every utterance. This suggests that the optimal modality for classification varies across utterances, and that selectively utilizing a single modality rather than applying fusion indiscriminately is more effective for improving performance. 

\subsection{Ablation study}

\subsubsection{Acoustic stage}
To investigate the impact of each method applied in the acoustic stage, we broke down and analyzed the overall performance in detail. For this purpose, we define four components as follows:

\begin{itemize} 
    \item \textbf{Fine-tuning}: The Whisper encoder, illustrated in Figure~\ref{fig:locality-aware}, was fine-tuned using the collected self-talk dataset.
    \item \textbf{Custom Positional Encoding (PE)}: The sinusoidal positional encoding, as shown in Figure~\ref{fig:locality-aware}, was replaced with a custom positional encoding inspired by a prior study~\cite{sousa2020dense}.
    The goal was to enable the encoder to learn that shorter temporal gaps between utterances correspond to fewer class transitions. To capture the temporal relationships, we concatenated multiple segments and fed them into the Whisper encoder. However, the original Whisper’s positional encoding only reflects the sequential order of audio frames by assigning a fixed index (0–1500) to each frame, without considering the actual temporal gaps between utterances. As a result, it cannot represent the timing or spacing of utterances that occur in actual situations. To overcome this limitation, we generated a new positional encoding based on the actual occurrence time of each segment within a session.
    \item \textbf{Static LAEA}: An EMA with a fixed weight was applied to the embeddings extracted from the encoder when the time gap between consecutive utterances was less than 4 seconds; the $\alpha$ parameter was set to 0.5 for LAEA illustrated in Figure~\ref{fig:locality-aware}.
    \item \textbf{Adaptive LAEA}: An adaptive, trainable EMA-based embedding adaptation was applied to the encoder outputs. (See Section~\ref{sec:laea})
  
\end{itemize}

\begin{table}[t]
\caption{Ablation study on acoustic stage}
\footnotesize
\vspace{-0.1in}
\centering
\begin{tabular}{c c| c c c c|c c c| c}
\toprule
& & \makecell{\textbf{Fine-} \\ \textbf{tuning}} & \makecell{\textbf{Custom} \\ \textbf{PE}} & \makecell{\textbf{Static} \\ \textbf{LAEA}} & \makecell{\textbf{Adaptive} \\ \textbf{LAEA}} &   \makecell{\textbf{Negative} \\ \textbf{self-talk}} & \makecell{\textbf{Positive} \\ \textbf{self-talk}} & \textbf{Others}  & \makecell{\textbf{Macro} \\ \textbf{Avg.}} \\
\midrule

& \textbf{Baseline 1} & - & - &  - & - & 0.64 & 0.15 & 0.80 & 0.53 \\
& \textbf{Baseline 2} & $\bigcirc$ & - & - & - & 0.78 & 0.62 & 0.87 & 0.75 \\
& \textbf{Baseline 3} & $\bigcirc$ & $\bigcirc$ & - & - & 0.65 & 0.43 & 0.80 & 0.63 \\
& \textbf{Baseline 4} & $\bigcirc$ & - & $\bigcirc$ & - & 0.74 & 0.61 & 0.84 & 0.73 \\
\rowcolor{gray!20} 
& \textbf{MutterMeter} & $\bigcirc$ & - & - & $\bigcirc$ & 0.81 & 0.71 & 0.89 & 0.80 \\

\bottomrule
\end{tabular}
\label{tab:acoustic_result}
\end{table}

To examine the contribution of each component in the acoustic stage, we designed four baseline models with different configurations. Baseline 1 consists of a pre-trained Whisper encoder followed by a classifier, with no additional components. Baseline 2 applies fine-tuning to the encoder using the collected self-talk dataset. 
Baseline 3 builds upon Baseline 2 by incorporating the custom PE, replacing the original sinusoidal PE.
Finally, Baseline 4 builds on Baseline 2 but employs the Static LAEA.

Table~\ref{tab:acoustic_result} shows that MutterMeter achieves the highest macro-averaged $F_{1}$ score among all baselines.
Notably, compared to the lowest-performing baseline 1, our method achieves a 0.27 increase in the macro-averaged $F_{1}$ score — corresponding to a 51\% improvement. Although the pre-trained Whisper model was trained on a large amount of data, it was originally trained for an ASR task. Thus, it was unable to properly extract embeddings for self-talk detection, leading to poor performance. In particular, positive self-talk shows a relatively low performance with an $F_1$ score of 0.15 compared to other classes. This is because, in terms of acoustic characteristics, encouraging or goal-directed positive self-talk often exhibits patterns similar to utterances in the others class, making it difficult for the model to distinguish between them.
Baseline 2 shows a substantial improvement compared to Baseline 1, with a 0.22 increase in the macro-averaged $F_{1}$ score, indicating that fine-tuning enabled the encoder to extract embeddings for self-talk detection. However, its performance for positive self-talk remains relatively low ($F_{1}$ score = 0.62).
Baseline 3 achieves a macro-averaged $F_{1}$ score of 0.63, showing lower performance than Baseline 2 and the LAEA-based approaches. We expected that the custom PE would enable the encoder to learn temporal relationships between utterances more effectively. However, our dataset size was insufficient to adequately train the large-scale parameters of the pre-trained Whisper model. Since Whisper is originally optimized for an architecture trained on large-scale speech datasets, modifying this structure by applying the custom PE may have limited the model from fully leveraging its pre-trained representational power.
Baseline 4 also shows lower performance than Baseline 2. This suggests that the Static EMA, by indiscriminately averaging embeddings between temporally adjacent utterances, distorted the embeddings that capture the unique characteristics of each utterance, leading to performance degradation.
MutterMeter, through Adaptive LAEA, adaptively adjusts the embeddings for each utterance, resulting in clearer class boundaries in the embedding space (Figure~\ref{fig:effect_embedding_adaptation}). As a result, compared to Baseline 2, which applies only fine-tuning, the macro-averaged $F_{1}$ score increased from 0.75 to 0.80 (a 6.7\% improvement), demonstrating an overall performance enhancement. In particular, positive self-talk shows the most pronounced improvement, with a 27.9\% increase, while negative self-talk also improved by 7.7\%.

\subsubsection{Linguistic stage}

To investigate how the strategies discussed in Section~\ref{linguistic_stage} affect the self-talk detection performance of the linguistic stage, we established the following baselines. 

\begin{itemize}
    \item \textbf{MutterMeter–w/o contextual dependency}: Transcription and subsequent classification were performed using the single-utterance transcription strategy described in Table~\ref{tab:transcription_quality}.
    \item \textbf{MutterMeter-w/o temporal aspects}: Transcription and subsequent classification were performed using the contextual transcription w/o temporal aspect strategy described in Table~\ref{tab:transcription_quality}.
    \item \textbf{MutterMeter-w/o quantity aspects}: Transcription and subsequent classification were performed using the contextual transcription w/o quantity aspect strategy described in Table~\ref{tab:transcription_quality}.
    \item \textbf{MutterMeter}: Transcription and subsequent classification were performed using the contextual transcription strategy described in Table~\ref{tab:transcription_quality}.
\end{itemize}

\begin{table}[t]
\caption{Ablation study on linguistic stage}
\footnotesize
\vspace{-0.1in}
\centering
\begin{tabular}{l |c c c| c}
\toprule
& \makecell{\textbf{Negative} \\ \textbf{self-talk}} & \makecell{\textbf{Positive} \\ \textbf{self-talk}} & \textbf{Others}  & \makecell{\textbf{Macro} \\ \textbf{Avg.}} \\
\midrule

\textbf{MutterMeter-w/o contextual dependency} & 0.69 & 0.48 & 0.81 & 0.66 \\
\textbf{MutterMeter-w/o temporal aspects} & 0.65 & 0.51 & 0.78 & 0.65 \\
\textbf{MutterMeter-w/o quantity aspects} & 0.69 & 0.56 & 0.80 & 0.68 \\
\rowcolor{gray!20} 
\textbf{MutterMeter} & 0.71 & 0.58 & 0.81 & 0.70 \\

\bottomrule
\end{tabular}
\label{tab:linguistic_result}
\end{table}
As shown in Table~\ref{tab:linguistic_result}, MutterMeter outperforms the other baselines in self-talk detection, achieving a macro-averaged $F_{1}$ score of 0.70. Overall, the performance is proportional to the transcription quality, as presented in Table~\ref{tab:context_aware_transcription}. In particular, compared to the MutterMeter-w/o contextual dependency, MutterMeter improves the macro-averaged $F_{1}$ score by 0.04. Interestingly, for positive self-talk, the $F_{1}$ score increases from 0.48 to 0.58, representing a relative improvement of 20.8\%, which corresponds to a 46.1\% increase in correctly classified samples. This improvement could be attributed to the presence of murmured or unclear utterances in positive self-talk, making it difficult to achieve accurate transcription when relying solely on a single utterance. For negative self-talk, the $F_{1}$ score increases from 0.69 to 0.71 (+0.02), with the number of correctly classified samples increasing by 6.4\%. The others class shows no significant change in $F_{1}$ score.

\subsubsection{Fusion stage}

\begin{table}[t]
\centering
\caption{Ablation study on fusion stage}~\label{tab:fusion_result}
\footnotesize
\vspace{-0.1in}
\begin{tabular}{l|ccc|c}
\toprule
 & \makecell{\textbf{Negative} \\ \textbf{self-talk}} & \makecell{\textbf{Positive} \\ \textbf{self-talk}} & \textbf{Others}  & \makecell{\textbf{Macro} \\ \textbf{Avg.}} \\

\midrule
\textbf{Static weight fusion} & 0.80 & 0.71 & 0.88 & 0.80 \\ 
\rowcolor{gray!20} 
\textbf{Adaptive weight fusion} & 0.81 & 0.75 & 0.89 & 0.82 \\

\bottomrule
\end{tabular}
\end{table}

To examine the effect of the adaptive weight fusion proposed in the fusion stage, we used the static weight fusion as a baseline, assigning an equal weight of 0.5 to each modality. As shown in Table~\ref{tab:fusion_result}, the adaptive weight fusion achieves higher performance than the static weight fusion, with an improvement of 0.02 in the macro-averaged $F_{1}$ score. Specifically, the $F_{1}$ score increases by 0.04 for positive self-talk and by 0.01 for negative self-talk, while the number of correctly classified samples increases by 13.9\% and 7.0\%, respectively.
Interestingly, the macro-averaged $F_{1}$ score of the static weight fusion is almost identical to that of the acoustic model presented in Table~\ref{tab:overall_performance}. This suggests that while static weight fusion can improve classification performance for certain utterances by leveraging multimodal information, it may become ineffective when acoustic features alone are insufficient to distinguish utterances or linguistic features are degraded due to transcription errors. In such cases, assigning equal weights may instead hinder correct classification.

\subsection{Comparison with alternative approaches}~\label{sec:alter_approach}
\begin{table}[t]
\centering
\caption{Comparison with other approaches}~\label{tab:others_performance}
\footnotesize
\vspace{-0.1in}
\begin{tabular}{l|ccc|c}
\toprule
  & \makecell{\textbf{Negative} \\ \textbf{self-talk}} & \makecell{\textbf{Positive} \\ \textbf{self-talk}} & \textbf{Others}  & \makecell{\textbf{Macro} \\ \textbf{Avg.}} \\

\midrule
\textbf{Emotion2Vec}~\cite{ma-etal-2024-emotion2vec} & 0.26 & 0.19 & 0.20 & 0.21 \\ 
\textbf{TweetNLP}~\cite{camacho-collados-etal-2022-tweetnlp} & 0.57 & 0.35 & 0.59 & 0.50 \\
\textbf{Gemini-Text-zero-shot} & 0.64 & 0.53 & 0.75 & 0.64  \\
\textbf{Gemini-Text-few-shot} & 0.69 & 0.58 & 0.79 & 0.69   \\
\textbf{Gemini-Multimodal-zero-shot} & 0.66 & 0.52 & 0.78 & 0.66 \\
\textbf{Gemini-Multimodal-few-shot} & 0.69 & 0.52 & 0.79 & 0.67  \\
\rowcolor{gray!20} 
\textbf{MutterMeter} & 0.84 & 0.77 & 0.91 & 0.84 \\
\bottomrule
\end{tabular}
\end{table}

To examine the potential of alternative approaches for self-talk detection, we also evaluated several pre-trained models that could be adapted for this task. Specifically, we included representative models from SER (Emotion2Vec~\footnote{\url{https://huggingface.co/emotion2vec/emotion2vec_plus_base}}) , sentiment analysis (TweetNLP~\footnote{\url{https://huggingface.co/cardiffnlp/twitter-xlm-roberta-base-sentiment-multilingual}}), and LLM-based (Gemini~\footnote{\url{https://cloud.google.com/vertex-ai/generative-ai/docs/models/gemini/2-0-flash}}) approaches, and assessed their self-talk detection performance using our dataset. 
Note that Emotion2Vec and TweetNLP define their classes differently from MutterMeter; thus, we performed a class mapping to align them with our label schema and evaluated their performance accordingly. For example, in Emotion2Vec, angry, disgusted, fearful, and sad were mapped to negative self-talk, happy and surprised were mapped to positive self-talk, and neutral, other, and unknown were mapped to others. In TweetNLP, negative was mapped to negative self-talk, positive to positive self-talk, and neutral to others.
In addition, several variants of LLM-based models were constructed depending on the input modality (text-only or multimodal audio-text) and prompting method (few-shot or zero-shot). A detailed example of the prompt used for Gemini is provided in Appendix~\ref{appendix_prompt}.

Table~\ref{tab:others_performance} shows that MutterMeter outperforms alternative approaches in self-talk detection. Specifically, as discussed in Section~\ref{sec:background_and_motivation}, SER and sentiment analysis generally struggle to accurately detect self-talk. 
For LLM-based approaches, the text-only few-shot prompting setting achieves a relatively higher macro-averaged $F_{1}$ score (0.69) than the zero-shot setting; however, this performance is only comparable to that of MutterMeter’s linguistic model in Table~\ref{tab:overall_performance}. 
In the multimodal input setting, despite the additional use of audio information, the performance does not improve and even shows a slight decline compared to the text-only setting. Furthermore, few-shot prompting exhibits no noticeable advantage over zero-shot prompting. These results suggest that LLMs do not effectively leverage the audio modality for self-talk detection. Specifically, although audio information is provided, the models appear unable to capture the subtle acoustic cues essential for detecting self-talk, as they were not inherently trained for this purpose.

\subsection{System cost}

\begin{table}[t]
\centering
\caption{Stage-wise latency and processing ratio}~\label{tab:latency_performance}
\footnotesize
\vspace{-0.1in}
\begin{tabular}{c|c|c|c}
\toprule
\textbf{Side} & \textbf{Stage} & \textbf{Average latency (ms)} & \textbf{Processing ratio} \\
\midrule
\multirow{2}{*}{\textbf{Mobile side}} & Preprocessing stage & 20.9 & - \\
 & Acoustic stage & 2015 & 61\% \\
\cline{1-4}
\multirow{2}{*}{\textbf{Server side}} & Linguistic stage & 4298 & 7\% \\
 & Fusion stage & 0.8 & 32\%\\
\bottomrule
\end{tabular}
\end{table}

To examine how effectively the hierarchical architecture of MutterMeter reduces system cost, we measured the latency and processing ratio at each stage, as summarized in Table~\ref{tab:latency_performance}. The latency of the preprocessing and acoustic stages was measured on a Galaxy S21 device, while the linguistic and fusion stages were measured on a TITAN RTX server. In the preprocessing stage, latency was measured using a custom processing configuration with 100 ms audio segments as the processing unit, resulting in an average latency of 20.9 ms per segment. The acoustic, linguistic, and fusion stages were examined in terms of average latency per utterance. The acoustic stage exhibits an average latency of 2015 ms, while on the server side, the linguistic and fusion stages exhibit average latencies of 4298 ms and 0.8 ms, respectively.
By analyzing the processing ratio at each stage, we found that approximately 61\% of utterances are classified at the acoustic stage, indicating that a substantial portion of processing is completed on the mobile side. Meanwhile, about 7\% of utterances are processed at the linguistic stage, and the remaining 32\% are processed at the fusion stage.

The hierarchical architecture of MutterMeter provides a significant advantage over a structure in which all utterances are processed sequentially through every stage until the final fusion stage. MutterMeter is designed to allow early exit at intermediate stages once sufficient confidence is achieved. In particular, the acoustic stage accounts for 61\% of all utterances and exhibits substantially lower latency compared to the subsequent linguistic and fusion stages. This allows the system to make rapid decisions without incurring additional computational or transmission costs. Specifically, when self-talk detection is performed without early exit, the average latency per utterance is 6335 ms; however, applying the hierarchical architecture with early exit reduces the average latency to 3713 ms, achieving a 41\% reduction. These results demonstrate that MutterMeter’s stage-wise early-exit mechanism not only significantly reduces overall system latency but also improves the accuracy of self-talk detection, as shown in Table~\ref{tab:overall_performance}.

%% file: sections/07.Related_Work.tex
\section{Related Work}

\subsection{Speech-based understanding}

The technology that enables machines to interpret the meaning and emotions embedded in human speech has been continuously explored. Such technology can play a crucial role in automatically inferring the speaker’s cognitive and emotional states. From a broader perspective, speech-based understanding approaches are conceptually aligned with self-talk detection tasks.

A representative technology in this domain is SER, which aims to identify emotions from speech. Traditionally, approaches that use acoustic features such as pitch and energy as model inputs have been widely explored~\cite{ververidis2006emotional, dellaert1996recognizing}. Recently, end-to-end learning methods have been studied, in which raw audio waveforms or spectrograms are directly fed into transformer-based models~\cite{wang2021novel, he2023multiple}. In addition, speech-based pre-trained models such as Wav2Vec, Whisper, and HuBERT have been fine-tuned to achieve high performance~\cite{pepino2021emotion, wang2021fine}. Furthermore, multi-modal approaches that integrate audio, text, or video information have also been investigated~\cite{cheng2024emotion, yang2022disentangled, girish2022interpretabilty}.
Spoken Language Understanding (SLU) is also closely related, as it aims to extract semantic structures from users’ speech. Conventionally, studies have employed a separated pipeline consisting of ASR and Natural Language Understanding (NLU), where the speech signal is first converted into text and then semantically interpreted~\cite{faruqui2022revisiting}. More recently, end-to-end architectures that integrate ASR and NLU have been actively explored to reduce error propagation and latency~\cite{serdyuk2018towards, radfar2020end}. In addition, Benazir et al.~\cite{benazir2024speech} have also proposed a method to improve computational efficiency through a caching mechanism.

Conventional SER and SLU models are typically trained on datasets collected from clear, emotionally explicit, and interaction-oriented speech, assuming high-volume and grammatically structured utterances as input. However, self-talk differs fundamentally, often involving spontaneous thought processes, low-volume or whispered speech, and incomplete or self-directed expressions, making it difficult for conventional models to detect or interpret effectively. In this study, we view self-talk as a continuous and cognitively driven behavior shaped by internal motivation and situational context. Accordingly, we propose a new approach that considers semantic continuity and contextual dependency, leveraging surrounding utterances. We believe that this approach makes a meaningful contribution to research on speech understanding.

\subsection{Mobile sensing and mental state}

With the widespread adoption of smartphones, many studies have explored ways to infer various mental states by extracting contextual and behavioral features from the built-in sensors (e.g., GPS, accelerometer, microphone, Bluetooth, Wi-Fi) and mobile usage patterns (e.g., call and text frequency, app usage duration) of mobile devices. Wang et al.~\cite{wang2014studentlife} and Boukhechba et al.~\cite{boukhechba2017monitoring} proposed methods to assess college students’ mental health status and academic performance, while Zakaria et al.~\cite{zakaria2019stressmon} and Nosakhare et al.~\cite{nosakhare2019probabilistic} aimed to estimate stress and depression levels. Similarly, Morshed et al.~\cite{morshed2019prediction} and Saha et al.~\cite{saha2017inferring} focused on predicting mood instability. Meanwhile, Cao et al.~\cite{cao2017deepmood} presented a method for detecting bipolar disorder using key press timing and hand movement data during typing.
In addition to smartphone-based passive sensing, recent advances in wearable technologies have enabled unobtrusive and continuous collection of physiological signals, such as PPG, EEG, ECG, and EOG, over long periods. These signals provide objective indicators of mental and cognitive conditions, enabling the inference of states such as mental fatigue~\cite{kodikara2025fatiguesense, shen2008eeg, karim2023eeg}, stress~\cite{lueken2017photoplethysmography, suzuki2015stress}, anxiety~\cite{sahu2024wearable}, and cognitive load~\cite{knierim2025advancing}. Leveraging such mobile sensing capabilities allows accurate, real-world assessment of mental states without continuous intervention from medical professionals, offering new opportunities for personalized everyday support.

We can narrow the scope to audio-based approaches, where various studies have attempted to detect mental states by utilizing users’ speech collected through microphones. Lu et al.~\cite{lu2012stresssense} proposed a method for stress detection through StressSense, Rachuri et al.~\cite{rachuri2010emotionsense} introduced EmotionSense for emotional state inference, and Salekin et al.~\cite{salekin2018weakly} presented an approach for detecting social anxiety and depression.
However, these approaches have mainly focused on the acoustic characteristics of speech and therefore have limitations in conducting in-the-wild analyses that consider speech content and speaking style~\cite{wang2023power}.
With recent advances in Natural Language Processing (NLP), several studies have attempted to extract linguistic features embedded in speech and combine them with acoustic features to infer mental states. Wang et al.~\cite{wang2023power} proposed a voice diary–based method to assess auditory verbal hallucination severity using recorded speech and automatic transcriptions. Similarly, Pillai et al.~\cite{pillai2024investigating} explored detecting mental disorders, such as suicidal ideation, from voice diary data.

This study builds on previous speech-based mental state inference research by using speech data from mobile devices. However, the key distinction of this work lies in focusing on self-talk, a form of spontaneous speech that naturally occurs in everyday life and has been largely overlooked in previous studies. 
From a technical perspective, this study proposes an integrated approach that combines acoustic and linguistic information to analyze spontaneous speech in real-world contexts. Unlike prior voice diary–based studies that focus on speech recorded in specific moments or controlled environments, our work explores the feasibility of analyzing speech in more naturalistic, everyday settings. We believe that self-talk detection can complement existing mental state inference methods, contributing to the development of a continuous, fine-grained mental health monitoring system that operates effectively in daily life.

%% file: sections/08.Conclusion.tex
\section{Conclusion}
In this work, we present MutterMeter, a novel mobile system that automatically detects self-talk by analyzing audio collected from earable microphones. MutterMeter leverages the temporal and contextual information of utterances and employs a hierarchical architecture to detect self-talk accurately and efficiently. We present extensive experiments using a dataset consisting of 31.1 hours of audio recordings collected from 25 participants in real tennis-playing situations, validating the performance of MutterMeter. 

We believe that MutterMeter holds strong potential to understand individuals’ psychological states through spontaneous language in daily life and to inspire subtle, everyday nudges that enhance self-awareness and emotional well-being. Moving forward, our future work will focus on broadening the applicability and scalability of MutterMeter.
Beyond self-talk detection, we aim to technically advance MutterMeter so that it can analyze why and under what contexts self-talk occurs, thereby providing individuals with meaningful feedback. While this study focused on tennis-playing situations where self-talk naturally occurs, future work will extend MutterMeter to diverse real-world contexts to evaluate its generalizability. Finally, optimizing MutterMeter for fully on-device operation could enhance privacy protection, given the personal and sensitive nature of self-talk.

%% file: sections/A.Appendix.tex
\section{Appendix: The Prompt Formats}~\label{appendix_prompt}

Table~\ref{tab:prompt_base_text} shows the zero-shot and few-shot prompts for the LLM that takes only text as input. To provide audio feature descriptions to the prompt, we converted the audio features into text using Table~\ref{tab:feature_ranges}. Table~\ref{tab:prompt_base_multi} shows the zero-shot and few-shot prompts for the LLM that takes both audio and text as input. In the few-shot case, nine examples were used as shown in Table~\ref{tab:prompt_shot}. For the audio–text LLM, each shot included the corresponding audio.

\begin{table}
\caption{Prompt configurations for text-based LLMs}~\label{tab:prompt_base_text}
\footnotesize
\begin{tabular}{p{2cm} | p{13.5cm}}
\textbf{Model} & \textbf{Prompt} \\
\midrule
\raggedright Gemini-Text-zero-shot & Classify the following tennis utterance, considering its **text**, **audio features**, and the **context provided by the previous 10 utterances**. The input utterance text and historical utterances may be in Korean. Choose one of these three categories:
\newline- **Negative Self-Talk**: Blaming oneself, expressing negative thoughts or negative exclamations.
\newline- **Positive Self-Talk**: Encouraging oneself or expressing positive thoughts.
\newline- **Others**: General conversation, non-meaningful sounds like breathing, racket noise, or background noise, counting scores (e.g., "Fifteen-love!", "Deuce!"), or in/out calls (e.g., "Out!", "Fault!").
\newline

**Audio Feature Descriptions:**
\newline- **Pitch Variance**: How much the pitch (voice fundamental frequency) changes. A higher variance can indicate strong emotion or excitement.
\newline- **Pitch Mean**: The average pitch of the voice.
\newline- **Duration**: The length of the utterance in seconds.
\newline- **Intensity Mean**: The average loudness of the voice.
\newline- **Pitch Range**: The span between the highest and lowest pitch in the utterance
\newline- **Intensity Range**: The span between the loudest and quietest parts of the utterance 
\newline

Do **not** explain your reasoning. Do **not** return any additional text, description, or commentary. Respond with **only** one of the classification labels.
\end{tabular}

\begin{tabular}{p{2cm} | p{13.5cm}}
\midrule
\raggedright Gemini-Text-few-shot & Classify the following tennis utterance, considering its **text**, **audio features**, and the **context provided by the previous 10 utterances**. The input utterance text and historical utterances may be in Korean. Choose one of these three categories:
\newline- **Negative Self-Talk**: Blaming oneself, expressing negative thoughts or negative exclamations.
\newline- **Positive Self-Talk**: Encouraging oneself or expressing positive thoughts.
\newline- **Others**: General conversation, non-meaningful sounds like breathing, racket noise, or background noise, counting scores (e.g., "Fifteen-love!", "Deuce!"), or in/out calls (e.g., "Out!", "Fault!").
\newline

**Audio Feature Descriptions:**
\newline- **Pitch Variance**: How much the pitch (voice fundamental frequency) changes. A higher variance can indicate strong emotion or excitement.
\newline- **Pitch Mean**: The average pitch of the voice.
\newline- **Duration**: The length of the utterance in seconds.
\newline- **Intensity Mean**: The average loudness of the voice.
\newline- **Pitch Range**: The span between the highest and lowest pitch in the utterance
\newline- **Intensity Range**: The span between the loudest and quietest parts of the utterance 
\newline

Here are a few examples:
\newline \{shots\}
\newline

Do **not** explain your reasoning. Do **not** return any additional text, description, or commentary. Respond with **only** one of the classification labels.
\end{tabular}
\end{table}

\begin{table}
\caption{Prompt configurations for audio–text LLMs}~\label{tab:prompt_base_multi}
\footnotesize
\begin{tabular}{p{2cm} | p{13.5cm}}
\midrule
\raggedright Gemini-Multimodal-zero-shot & Classify the following tennis utterance, considering its **text**, **audio features**, and the **context provided by the previous 10 utterances**. The input utterance text and historical utterances may be in Korean. Choose one of these three categories:
\newline- **Negative Self-Talk**: Blaming oneself, expressing negative thoughts or negative exclamations.
\newline- **Positive Self-Talk**: Encouraging oneself or expressing positive thoughts.
\newline- **Others**: General conversation, non-meaningful sounds like breathing, racket noise, or background noise, counting scores (e.g., "Fifteen-love!", "Deuce!"), or in/out calls (e.g., "Out!", "Fault!").
\newline

Do **not** explain your reasoning. Do **not** return any additional text, description, or commentary. Respond with **only** one of the classification labels.

\end{tabular}

\begin{tabular}{p{2cm} | p{13.5cm}}
\midrule
\raggedright Gemini-Multimodal-few-shot & Classify the following tennis utterance, considering its **text**, **audio features**, and the **context provided by the previous 10 utterances**. The input utterance text and historical utterances may be in Korean. Choose one of these three categories:
\newline- **Negative Self-Talk**: Blaming oneself, expressing negative thoughts or negative exclamations.
\newline- **Positive Self-Talk**: Encouraging oneself or expressing positive thoughts.
\newline- **Others**: General conversation, non-meaningful sounds like breathing, racket noise, or background noise, counting scores (e.g., "Fifteen-love!", "Deuce!"), or in/out calls (e.g., "Out!", "Fault!").
\newline

Here are a few examples:
\newline \{shots\}
\newline

Do **not** explain your reasoning. Do **not** return any additional text, description, or commentary. Respond with **only** one of the classification labels.
Here are a few examples:
\end{tabular}
\end{table}

\begin{table}
\caption{Few-shot prompt format and example}~\label{tab:prompt_shot}
\footnotesize
\begin{tabular}{p{2cm} | p{13.5cm}}
\midrule
\raggedright Format & **Previous Utterances History (from oldest to newest, up to 10 utterances, comma-separated):**
\newline \{history\}
\newline **Current Utterance to Classify:**
\newline Utterance: \{utterance\}
\newline Features: Pitch Variance=\{pitch\_variance\}, Pitch Mean=\{pitch\_mean\}, Duration=\{duration (sec)\} Intensity Mean=\{intensity\_mean\}, Pitch Range=\{pitch\_range\}, Intensity Range=\{intensity\_range\}, Pitch Contour=\{pitch\_contour\}.
\newline Classification: \{result\}
\end{tabular}

\begin{tabular}{p{2cm} | p{13.5cm}}
\midrule
\raggedright Example & **Previous Utterances History (from oldest to newest, up to 10 utterances, comma-separated):**
\newline " 여기 여기인가?", " 에", " 자신있게", " 이건 아니야", " 왜이러냐 왜이러냐 왜이러냐", " 아 배가아퍼", " 아 내 좋아하는 공이었는데", " 엥?", " 아 서버가 너무 안된다 진짜", " 자자자자"
\newline **Current Utterance to Classify:**
\newline Utterance:  자신있게 자신있게
\newline Features: Pitch Variance=High, Pitch Mean=Low, Duration=0.928813 (sec), Intensity Mean=High, Pitch Range=Wide, Intensity Range=Wide, Pitch Contour=The pitch suddenly rises at a certain moment.
\newline Classification: Positive Self-Talk
\end{tabular}
\end{table}

\begin{table}[t]
\centering
\caption{Feature value ranges used for prompt generation.}
\label{tab:feature_ranges}
\footnotesize
\begin{tabular}{p{2cm} | p{13.5cm}}
\toprule
\textbf{Feature} & \textbf{Selected Range or Description} \\
\midrule
\texttt{pitch\_variance} & Divided into tertiles (\emph{Low}, \emph{Midium}, \emph{High}) based on overall pitch variability. Represents stability or fluctuation in voice pitch. \\[4pt]

\texttt{pitch\_mean} & Categorized into tertiles (\emph{Low}, \emph{Midium}, \emph{High}) according to average fundamental frequency (F0). Reflects the speaker’s general pitch height. \\[4pt]

\texttt{duration} & Continuous value in seconds (\(t = \text{len(audio)} / sr\)). Used to capture the utterance length for prompt description. \\[4pt]

\texttt{intensity\_mean} & Divided into tertiles (\emph{Low}, \emph{Midium}, \emph{High}) representing average signal energy (proxy for loudness). \\[4pt]

\texttt{pitch\_range} & Computed as \(\max(F0) - \min(F0)\); further divided into tertiles (\emph{Wide}, \emph{Midium}, \emph{Narrow}) to indicate variation span of pitch. \\[4pt]

\texttt{intensity\_range} & Computed as \(\max(RMS) - \min(RMS)\); divided into tertiles (\emph{Wide}, \emph{Midium}, \emph{Narrow}) to indicate the dynamic range of loudness. \\[4pt]

\texttt{pitch\_contour} &
Derived from the temporal dynamics of fundamental frequency (F0) using several general conditions:  
\newline (1) If pitch variation is minimal, \emph{the pitch remains steady with almost no variation.}  
\newline (2) If there is a sudden, large upward change, \emph{the pitch suddenly rises at a certain moment.}  
\newline (3) If there is a sudden, large downward change, \emph{the pitch suddenly drops at a certain moment.}
\newline (4) If the pitch gradually increases over time, \emph{the pitch gradually tends to rise.}
\newline (5) If the pitch gradually decreases over time, \emph{the pitch gradually tends to fall.}
\newline (6) If the pitch fluctuates frequently, \emph{the pitch rises and falls repeatedly.}
\newline Otherwise, \emph{the pitch changes in a complex manner.} \\[4pt]

\bottomrule
\end{tabular}
\end{table}